\documentclass[pra,twocolumn,showpacs]{revtex4-2}

\usepackage[utf8]{inputenc}
\usepackage[T1]{fontenc}
\usepackage[english]{babel}

\usepackage{amsmath,amsthm,amssymb,dsfont,amscd,mathrsfs}
\usepackage[centercolon=true]{mathtools}
\usepackage{graphicx,microtype,slashed,bigints}
\usepackage[usenames,dvipsnames]{xcolor}
\usepackage{simplewick,xspace}

\usepackage{hyperref}
\usepackage[capitalize]{cleveref}
\hypersetup{
  linkcolor=red,
  colorlinks=true
}

\theoremstyle{plain}

\theoremstyle{definition}

\newcommand*{\tp}[1]{_{\textup{#1}}}

\newcommand*{\IN}{\tp{in}}

\newcommand*{\AC}{\tp{ac}}

\newcommand*{\swap}[2]{\let\temp#1 \let#1#2 \let#2\temp \let\temp\relax}
\swap{\epsilon}{\varepsilon}
\swap{\theta}{\vartheta}
\swap{\phi}{\varphi}

\DeclareMathOperator{\Swap}{Swap}
\DeclareMathOperator{\supp}{supp}

\DeclareMathOperator*{\slim}{s-lim}

\DeclarePairedDelimiter{\norm}{\lVert}{\rVert}

\newcommand*{\deq}{\coloneqq}

\newcommand*\cg[1]{
   \vbox{%
     \hrule height 0.5pt
     \kern0.25ex
     \hbox{%
       \kern-0.1em
       \ifmmode#1\else\ensuremath{#1}\fi
       \kern-0.1em
     }
   }
}

\newcommand{\bra}[1]{\langle#1|}
\newcommand{\ket}[1]{|#1\rangle}
\newcommand{\braket}[2]{\langle#1|#2\rangle}
\newcommand{\ketbra}[2]{{\ket{#1}\bra{#2}}}

\newcommand*{\dd}{\mathrm{d}\mkern0mu}

\renewcommand*{\H}{\mathcal H}

\newcommand*{\ST}{\mathsf{T}}

\DeclareMathOperator{\Ran}{Ran}

\begin{document}
\title{Scattering and perturbation theory for discrete-time dynamics}

 \author{Alessandro \surname{Bisio}}
\email[]{alessandro.bisio@unipv.it}
\affiliation{Dipartimento di Fisica, Universit\`a di Pavia, via Bassi 6, 27100 Pavia, Italy}
\affiliation{Istituto Nazionale di
  Fisica  Nucleare, Sezione di Pavia, Italy}

\author{Nicola \surname{Mosco}} \email[]{nicola.mosco@unipv.it}
\affiliation{Dipartimento di Fisica, Universit\`a di Pavia, via Bassi 6, 27100 Pavia, Italy}
\affiliation{Istituto Nazionale di Fisica Nucleare, Sezione
  di Pavia, Italy}

\author{Paolo \surname{Perinotti}}\email[]{paolo.perinotti@unipv.it}
\affiliation{Dipartimento di Fisica, Universit\`a di Pavia, via Bassi 6, 27100 Pavia, Italy}
\affiliation{Istituto Nazionale
  di Fisica Nucleare, Sezione di Pavia, Italy}

\begin{abstract}
  We present a systematic treatment of scattering processes
  for quantum systems whose time evolution is discrete.
 We define and show some general properties of the scattering
 operator, in particular the conservation of quasi-energy
 which is defined only modulo $2 \pi$.
 Then we develop two perturbative techniques for the power
 series expansion of the scattering operator, the first one
 analogous to the iterative solution of the Lippmann-Schwinger equation, the
second one to the Dyson series of perturbative
Quantum Field Theory.
We use this formalism to compare the scattering
amplitudes of a continuous-time model and of the
corresponding discretized one. We give a rigorous
assessment of the comparison for the case of bounded free Hamiltonian, as in
a lattice theory with a bounded number of
particles.
Our framework can be applied to a wide class of
quantum simulators, like quantum walks and quantum
cellular automata.  As a case study, we analyse the
scattering properties of a one-dimensional
cellular automaton with locally interacting
fermions.

\end{abstract}

\maketitle

\emph{Introduction.} Simulation of the time evolution of an arbitrary quantum system on a
classical computer is a computationally hard task. In one of the seminal papers of quantum 
computation~\cite{Feynman:1982ab}, Feynman suggested that this problem could be evaded by
having a quantum system simulate another one.
This intuition was proved to be correct~\cite{Lloyd1073}: a quantum
computer (or universal quantum simulator)
can indeed efficiently simulate any quantum system evolving through local interactions.

In recent years, there has been an increasing interest in quantum
simulators, with a broad range of theoretical proposals (see
e.g.~Refs.~\cite{PhysRevA.66.012305,Berry2007,PhysRevLett.102.180501,
PhysRevLett.106.170501,jordan2012quantum,Notarnicola2015,PhysRevLett.114.090502,PhysRevA.95.023604}
and the reviews~\cite{Cirac:2012aa,RevModPhys.86.153}), as well as various
experimental proofs of concept~\cite{Weimer2010382,Barends:2015aa,Hartmann_2016,Gross995,Martinez2016516}.

Except for  selected cases where the dynamics of the system can be mapped into
the dynamics of the simulator (e.g.~in the trapped ion simulation of the
Dirac Equation~\cite{Gerritsma:2010ab}), one has to
engineer a discrete quantum model of the system to be simulated.
The most studied class of quantum simulators (as the ones which we previously
cited)
are ``Hamiltonian-based''. In this setting, the experimenter is
supposed to be able to turn on
and turn off Hamiltonians from a given set.
Another class of quantum simulators
are Quantum Cellular Automata (QCA)
(see
Refs.~\cite{Succi:1993aa,Meyer:1996aa,PhysRevE.57.54,Schumacher:2004ab,ARRIGHI2011372,Gross:2012ab,freedman2019classification}
and the recent reviews~\cite{Arrighi:2019aa,farrelly2020review})
which consist of a
translational invariant
network of local quantum gates implementing the discrete-time evolution of a
lattice of quantum systems, each one interacting only with a
finite number of neighbours.
QCAs and in particular Quantum Walks (QWs)~\cite{venegas2012quantum}, which can be thought of as the one
particle sectors of QCAs, have been considered as a simulation tool for
relativistic quantum fields~\cite{Bialynicki-Birula:1994ab,DiMolfetta2013,Arrighi:2013ab,Farrelly:2014ac,PhysRevA.93.052301, DAriano:2014ae,arrighi2019quantum}  and as discrete approaches for studying the
foundations of Quantum Field
Theory~\cite{DAriano:2014ae,Bisio:2015aa,bisio2016quantum,bisio2016special,PhysRevA.100.012105}.

Once a simulation framework has been chosen, one needs some
tools to assess the quality of the simulation and quantify
how close the evolution of the discrete model and that of
the target system are.  This is usually achieved by applying
the Lie-Trotter-Kato product formula~\cite{hall2013quantum}
as shown in Ref.~\cite{Lloyd1073} for a quantum system whose
Hamiltonian is the sum of local interactions, 
i.e.~$H = \sum_{i=1}^k H_i$.  Then, the discrete time-step
evolution
$(e^{i H_1 \tau } \dots e^{i H_k \tau})^{t/\tau}$\footnote{This is
  the first order Lie-Trotter-Kato formula.  Higher-order
  approximants, that could be more efficient, have also been
  considered~\cite{SUZUKI1992387,PhysRevA.91.052327}}
approaches the target evolution $e^{i H t}$ as
$\tau \to 0$
\footnote{If the system to be simulated is a
  continuous theory,
the error due to space
discretisation should also be quantified.
In this regard, we point out the following recent analysis on
  the interplay between the continuous-time limit and
  continuous space limit~\cite{DiMolfetta2019}.}.  Typically,
if a fixed error threshold must not be exceeded, the larger
$t$ is, the smaller $\tau$ has to be: long time evolutions
demand quantum simulators which evolve over very short time
steps. For finite dimensional systems, a detailed
account of the errors for quantum simulations based
on product formulas can be given\cite{PhysRevLett.123.050503,PhysRevX.11.011020}.

What happens now if we consider the limit
$t\to \infty$?  This is the situation we encounter
in scattering processes
\cite{reed1979iii,newton2013scattering} in which
we study the evolution of wave-packets whose
evolution is asymptotically free in the far past
and in the far future. The scattering operator
(or $S$-matrix) is a key source of information
about a physical system, and hence the importance of
understanding it in quantum
simulators with a
discrete time evolution.  Since an $S$-matrix can
be defined only for infinite dimensional systems,
the methods employed for finite dimensional
systems are no longer of use.  Moreover, we cannot
even generally claim that the scattering operator
of a continuous-time theory is the limit of the
scattering operator of a Trotterized model. This
would be true if the two limits $\tau \to 0$ (for the
convergence of the Lie-Trotter-Kato formula) and
$t\to \infty$ (inherent in the definition of the $S$-matrix)
commuted. From the Moore-Osgood theorem, a
sufficient condition for the exchange of two
limits is the uniform convergence of one of the
two. Unfortunately, the limit of
the Lie-Trotter-Kato formula is generally only
locally uniform in $t$\cite{chernoff1974product}.

One could bypass this technical issues by
considering a finite dimensional simulation that
runs for a sufficiently large time $T$ such that
the scattering process takes place.  This is a
standard procedure and it is the one implemented
by Ref.\cite{jordan2012quantum}, which addresses
the problem of simulating scattering amplitudes in
quantum field theory.  This approach has the
technical advantage that the problem can analysed
in a more manageable finite-dimensional framework
and the mathematical tools of, for example,
Ref. \cite{childs2019theory} can be applied in
order to asses the error.  On the other hand, this
procedure is not optimized for scattering
processes since it requires that any process, not
only the scattering ones, is simulated within the same
error threshold.
We may expect that, if we 
focus on scattering amplitudes (and maybe only a
subset of them), then the time step $\tau$
(required to obtain the same fidelity) could be
smaller.  A parallel reasoning has been made by
in
Ref.\cite{Heyleaau8342}, where the error introduced by
Lie-Trotter-Kato formula has been evaluated for
localized observables.

Therefore, we are asking the following question: how small
$\tau$ has to be such that the scattering
amplitudes of the discrete-time simulation
reliably recover the scattering amplitudes of the
continuous model?

In this letter we make the first fundamental steps
to answering such a question.
We develop the theoretical tools needed to
define and compute the scattering operator for a
for quantum systems whose time evolution is
discrete. 
We will show how to adapt some of the
tools of standard continuous time scattering
theory to the discrete time case.  In particular,
we two approaches to the perturbative expansion for the
$S$-matrix: the first one is
analogous to the expansion of the
Lippmann-Schwinger equation, the second one to the
Dyson series of perturbative Quantum Field Theory.
Our analysis will show that time discreteness can
introduce a richer diversity of scattering
phenomena. This is an analogous of the Umklapp
scattering in solid state physics and it is due to
the fact that, for discrete time evolution, the
quasi-energy is conserved modulo a constant.

We then discuss the comparison between the scattering amplitudes of a continuous-time theory and 
the ones of a corresponding discrete-time one.
For systems with bounded energy, we 
prove and quantify the convergence of the
discrete-time scattering amplitudes to the
continuos time one as the discretization steps
$\tau$ goes to $0$.

Finally, we apply the theoretical analysis to a
one-dimensional discrete model for which the
two-particle dynamics is analytically known.

\emph{General scattering theory for discrete-time dynamics.}
Let us assume that a single time-step evolution is described by the unitary
operator $U \in \mathcal{B}(\H)$ which may describe many
particles in interaction or one particle with a potential. We denote
with  $U_0 \in \mathcal{B}(\H)$ the
 corresponding free evolution, i.e.~the evolution of our quantum
 system when the interaction (or the potential) is neglected.
 As usual, we assume that
$U_0$ is ``easy'', in the sense that it can be fully
diagonalised (e.g.~$U_0 = e^{-i H_0}$ with $H_0 =
\frac{\hbar^2}{2m} \nabla^2$,
the Hamiltonian of a non-relativistic  free particle).
In this paper we will consider elastic scattering
processes, the generalisation to multichannel scattering being left
for future works.
 The main issue in a theory of scattering is to give a precise
meaning to the statement ``$U^n\ket\psi$ looks asymptotically free as $n
\to -\infty$''. For that statement to be true, there must be a
state $\ket{\psi}\IN \in \H$ such that
  $\lim_{n\to -\infty} \norm{U_0^n\ket{\psi}\IN - U^n\ket{\psi}\IN} = 0.$
An analogous statement can be made for the future
asymptotic regime ($n \to +\infty$).
Therefore, one requires that the wave operators $ \Omega_\pm$
\begin{align}
  \Omega_\pm \deq \slim_{n\to \mp\infty}  {U^{\dag n}} {U_0}^n P\AC(U_0),
\end{align}
exist~\cite{reed1979iii} ($\slim$ is the limit in the
strong operator topology,
  and $P\AC(U_0)$ is the
projector on the subspace of absolute continuity of $U_0$).
  As it is the 
case in most applications, we will assume that $U_0$ has only absolute
continuous spectrum and has a generalised eigenvector
expansion of the kind
\begin{align}
  \label{eq:1}
  U_0 = \int_{B}\!\! e^{-i \omega(k)} \ketbra{k}{k} dk,
\end{align}
where $B \subseteq \mathbb{R}^n$, $\omega(k)$ is smooth, and
we employed the Dirac notation for generalised eigenvectors
\footnote{
  Including internal finite degrees of
  freedom, e.g.~spin, is a rather
  straightforward generalisation, which we omit in order to avoid a
  cumbersome notation.
}.
For example, if $U_0$ describes the evolution of a free particle,
then $k$ denotes the momentum and $B = \mathbb{R}^3$.
On the other hand, if the particle evolves on a discrete lattice
then $B \subseteq \mathbb{R}^n$ is the first Brillouin zone
and $k$ is the quasi-momentum.

The existence of the wave operators is a non-trivial and central
problem in scattering theory.
Ref.~\cite{Suzuki2016} proves
a nice generalisation of the Kato-Rosenblum theorem for unitaries
which states that, if $U-U_0$ is a trace-class operator, then $\Omega_{\pm}$
exist and $ \Ran(\Omega_+)=\Ran(\Omega_-) = \Ran(P\AC(U)) $.
If $\Omega_{\pm}$ exist, it is easy to see that they are isometric
($\Omega_{\pm}^\dag \Omega_{\pm} =I$) and obey the intertwining
relation
\begin{align}
  \label{eq:2}
  U \Omega_{\pm} = \Omega_{\pm}U_0.
\end{align}
Moreover, if $\Omega_{\pm}$ have the same range,
$\Ran(\Omega_+)=\Ran(\Omega_-)$, the scattering operator (or $S$-matrix)
\begin{align}
  \label{eq:3}
  S\coloneqq \Omega_{-}^\dag \Omega_+,
\end{align}
is a unitary operator. The $S$-matrix
 is the operator that relates \emph{in} and \emph{out} asymptotes, i.e.~incoming and
 outgoing particles,
and it is the main object of scattering
theory.

A straightforward consequence of Eq.~\eqref{eq:2} is the following commutation relation
\begin{align}
  \label{eq:4}
  [S,U_0] =0.
\end{align}
Despite its simplicity, Eq.~\eqref{eq:4} has important consequences.
The analogous equation for a continuous time dynamics would have been
$ [S, e^{-i H_0 t}] =0$ for any $t\in \mathbb{R}$, which implies $ [S, H_0 ] =0$: namely,
scattering processes conserve the energy.
However, there exist more than one exponential
representation of a unitary operator, for example we may have
$U_0 = e^{-i H_0 }$ with $H_0 = \int_{\mathbb{R}} p^2 \dd E_p$
or
$U_0 = e^{-i \tilde{H}_0 }$ with $\tilde{H}_0 = \int_{\mathbb{R}} (p^2
\mod 2\pi) \, \dd E_p$:
the energy eigenstates whose corresponding eigenvalues differ by $2
\pi$ are identified. Therefore, from Eq.~\eqref{eq:4}
we may only infer that (quasi-)energy  is conserved ``modulo $2 \pi$'' (in the
characteristic units of the discretised model).
This feature is also present in quantum systems with a
time-period driving~\cite{PhysRevA.91.033601}.
Such a periodicity in the energy conservation is responsible for a richer
scattering phenomenology in discrete time models.
This effect bears analogies with the Umklapp scattering in solid state
physics which is caused by the periodicity in momentum space due to
space discretisation.
In the following we will further discuss this
feature with the help of an explicit example.

\emph{Perturbative methods: the Lippmann-Schwinger equation.}
Once we know that the scattering operator of our dynamical
model is well defined, we need techniques that allow us to
compute the probability amplitude of scattering processes.
In the continuous-time framework,
these tools can be provided by an iterative solution of  the
Lippmann-Schwinger equation within
time-independent perturbation theory.
We now show the analogous of this perturbative method
for the discrete-time case.

As in the continuous case, the starting point is the
assumption that the expression $\Omega_\pm \ket{k}$,
suitably interpreted, is well defined \footnote{To determine
  whether this is possible is highly non trivial task and it
  is usually adressed case by case (see Chapter XI.6 of
  Ref.\cite{reed1979iii})}.  Then, we can show
\cite{supplement} that the improper matrix elements of the
scattering matrix are given by the following
equation:
\begin{align}
  \label{eq:5}
&  \begin{aligned}
      \bra{k'}S-I\ket{k} =& \lim_{\epsilon \to 0^+} 2 \pi \,
      \delta_{2 \pi}(\omega(k')-\omega(k)) \\
\times &\bra{k'} T(e^{-i\omega(k) + \epsilon})
  \ket{k},
  \end{aligned}\\
 & \nonumber
  T(z) \coloneqq  W +W G(z) U_0 W,
  \quad
W\coloneqq  U^\dag_0  U-I,
\end{align}
where  $G(z) = (zI - U)^{-1}$ is the resolvent of
  $U$ and $ \delta_{2 \pi}(x)$ denotes the Dirac comb
  with period equal to $2 \pi$.
The operator $G(z)$ obeys the  Lippmann-Schwinger equation
$
  G(z) = G_0(z) + G_0(z)(U-U_0)  G(z)
$
(where $G_0(z) \coloneqq (zI - U_0)^{-1}$) which yelds
to the following Lippmann-Schwinger equation for
$T(z)$
\begin{align}
  \label{eq:10}
  T(z) =  W +W G_0(z) U_0  T(z),
\end{align}
whose solution can be formally given in terms of the following
Born series
\begin{align}
  \label{eq:12}
  \begin{aligned}
     T(z) =  \sum_{n=0}^{\infty}(W G_0(z) U_0)^nW.
  \end{aligned}
\end{align}
By substituting Equation~\eqref{eq:12} in Equation~\eqref{eq:5}
we obtain a series expansion for the matrix elements of $S$
\begin{align}
  \label{eq:13}
     &\bra{k'}S-I \ket{k} = 2 \pi
     \delta_{2 \pi } (\omega(k')- \omega(k))\cdot \\
   &  \lim_{\epsilon \to 0^+}
  (
\bra{k'} W
  \ket{k} + \bra{k'} WG_0 (e^{-i\omega(k)+\epsilon}) U_0
W
\ket{k} +
\dots ). \nonumber
\end{align}
The convergence of the Born series depends on the existence
of the inverse of the operator
$I - (U_0^\dag U-I)G_0(z) U_0$ and therefore on the spectral
radius of $(U_0^\dag U-I)G_0(z)$. Let us consider the
simplest case in which
$ U = U_0V_\chi$ is the product of a free evolution $ U_0$
and
$V_\chi \coloneqq \sum_x e^{-i \chi f(x)} \ketbra{x}{x} )$ with
$f(x) \neq 0$  only on a finite set.
Then $(U_0^\dag U-I)$ is of finite rank and the convergence of
the Born series can be easily established. In particular, it
always
converges for sufficiently small $\chi$.

\emph{Discrete-time scattering vs continuous-time
  scattering} The formalism of the previous
section allows us to address comparison between
the scattering amplitudes of a continuous-time
theory described by a Hamiltonian $H := H_0 + V$,
with a bounded potential $|V|< +\infty$,
and the scattering amplitudes of the discretized
theory $U := e^{-iH_0 \tau}e^{-iV\tau}$ ($\tau$ is
the size of the temporal step).

Let us remind that, if we denote with $S^{c}$ the scattering
operator of the continuous theory, we have
 $\bra{k'}S^{(c)}-I\ket{k} = -2 \pi i\,
     \delta(
                         \omega_{k'}-\omega_k)
                        \bra{k'} T^{(c)}(\omega_k + i \epsilon)
                        \ket{k}$,
                        where $ T^{(c)}$ obeys
                        $T^{(c)}(z) =
  V + V G^{(c)}_0(z) T^{(c)}(z)$ with 
  $G^{(c)}_0(z) := (z-H_0)^{-1}$. On the other
  hand, $S^{(\tau)}$ will denote the scattering
  operator of the discrete theory:
  $\bra{k'}S^{(\tau)}-I\ket{k} = -2 \pi i
    \delta_{\tfrac{2 \pi}{\tau}}(
                         \omega_{k'}-\omega_k) 
\bra{k'} {T}^{(\tau)}(e^{-i(\omega_k + i\epsilon)\tau})
  \ket{k}$, where ${T}^{(\tau)}(z) :=
  \tfrac{i}{\tau}T(z)$ and $T(z)$ was defined in Equation \eqref{eq:5}.
Let us make the
following assumptions: $i)$ the spectrum of $H_0$
is upper bounded by $\omega_M$; this condition
applies to a lattice theory with a bounded number
of particles with $\omega_M:= N\omega_{max} $
where $\omega_{max}:= \max_{k\in
  \mathsf{B}}\omega_k$ and $N$ is the maximum
number of particles.  $ii)$
$|G^{(c)}_0(\omega_k + i \epsilon) V|=\gamma <1$;
this technical assumption
guarantees the existence of $(I-G^{(c)}_0 V)^{-1}$ and
convergence of the Born series.
Then we can prove \cite{supplement}  that,
\begin{align}
  \label{eq:9}
\tau \leq \min
  (\tfrac{\sqrt{2-\gamma}-1}{|V|} ,
  \tfrac{\pi}{\omega_M} ) \implies
  T^{(\tau)}-T^{(c)}=\tau R(\tau) 
\end{align}
where the operator valued function $ R(\tau)$  is
holomorphic in $\tau$ and bounded.
Equation \eqref{eq:9} rigorously proves the
convergence
 $S^{(\tau)}\xrightarrow{\tau \to 0}
    S^{(c)}$.
At  the leading
order in $\tau$ we have
\begin{align}
  \begin{aligned}
    \bra{k'}S^{(\tau)}-&S^{(c)}\ket{k}  =- 2 \pi \,
    \delta(
                         \omega_{k'}-\omega_k
                         ) \cdot \\
  &                       
\cdot \tau \,\bra{k'} {T}^{(c)}(\omega_k + i\epsilon)V
    \ket{k}+O(\tau^2).   
  \end{aligned}\label{eq:expanddifferencescattering}
\end{align}
By further expanding $ {T}^{(c)}(\omega_k + i\epsilon)$
in Equation \eqref{eq:expanddifferencescattering}
as a function of $V$ we obtain
$\bra{k'}S^{(\tau)}-S^{(c)}\ket{k}  =- 2 \pi \tau \,
    \delta(
                         \omega_{k'}-\omega_k
                         ) \bra{k'}V^2
    \ket{k}+O(\tau |V|^3)+O(\tau^2)$
    which shows that the first deviation
    introduced by the discretization are quadratic
    in the potential.

    The bound for $\tau$ in Equation~\eqref{eq:9}
    quantifies the intuition that a larger energy
    band and a stronger potential demands smaller
    time steps if we want that the scattering
    amplitudes of the dicrete model and the ones
    of the continuous model are close to each
    other.  In particular, the condition
    $\omega_M\tau \leq 2 \pi$, guarantees that
    $\delta_{\tfrac{2
        \pi}{\tau}}(\omega_{k'}-\omega_k) =
    \delta(\omega_{k'}-\omega_k)$ and scattering
    processes between states with different energy
    values are suppressed. This condition is
    necessary for the convergence
    $S^{(\tau)}\xrightarrow{\tau\to 0}
    S^{(c)}$. If the spectrum of the
    energy is not bounded, i.e.
    $\omega_M = +\infty$ then the suppression of
    scattering processes at different energy is
    possible only if $\bra{k'} {T}^{(\tau)}
    (e^{-i(\omega_k + i\epsilon)\tau})
    \ket{k} \to 0$ as $\tau \to 0$ for any $k,k'$
    such that $\omega_k =
    \omega_{k'} + \tfrac{2\pi n}{\tau} $, $n\in \mathbb{Z}$.
    This depends on the model at hand and
   such an analysis is beyond the scope of
   the present paper.
   However,  one has that at the leading order ${T}^{(\tau)}
    (e^{-i(\omega_k + i\epsilon)\tau}) = V +
    O(V^2)$
    the condition is satisfied provided that the Fourier
    transform of the potential $\hat{V}(k,k')$
    decays sufficiently rapidly as
    $|k-k'|\to+\infty$.
    For example, in the limit case of $V(x) = \delta(x)$ this
    condition cannot be satisfied and the
    convergence $S^{(\tau)}\xrightarrow{\tau \to 0}
    S^{(c)}$ is not achieved.


\emph{Perturbative methods: interaction picture and Dyson's formula.}
A typical situation is the one in which the single-step unitary evolution has the
form
\begin{align}
  \label{eq:14}
  \begin{aligned}
    U \coloneqq   U_0U_{\rm int} = e^{-i H_0}e^{-i  \chi H_{\rm int}} ,\\
    U_0 \coloneqq e^{-i H_0}, \qquad U_{\rm int} \coloneqq e^{-i  \chi H_{\rm int}} ,
  \end{aligned}
 \end{align}
where $H_0$ denotes a free evolution Hamiltonian,
$H_{\rm int}$ is an interaction Hamiltonian and $\chi$ is a coupling
constant (both $H_0$ and $H_{\rm int}$ are assumed to be time-independent).
This is the situation one encounters  (up to a scale factor) when simulating an Hamiltonian
of the kind $H_0 + \lambda H_{\rm int}$ by alternating one step of a
free evolution and one step of interaction.
In this case, it is convenient to represent the dynamics in the
interaction picture as follows:
\begin{align}
  \label{eq:15}
  \begin{aligned}
    & \ket{\psi(t)}_I \coloneqq  U_0^{\dag t} \ket{\psi(t)}_S,
  & O_I(t) \coloneqq  U_0^{\dag t}  O_S  U_0^t, \\
  & \ket{\psi(t+1)}_S = U \ket{\psi(t)}_S,
  & O_S \deq O_S(t) = O_S(0),
  \end{aligned}
\end{align}
where $\ket{\psi(t)}$ is a generic state, $O$ is a generic
operator, and the subscripts $S$ and $I$ denote the Schrödinger
and interaction picture respectively.
From Equation~\eqref{eq:15} the time evolution in the interaction
picture is easily derived:
\begin{align}
  \label{eq:16}
  \begin{aligned}
    &\ket{\psi(t+1)}_I = U_{I}(t) \ket{\psi(t)}_I, \\
    &U_{I}(t) \coloneqq U_0^{\dag t} U_{\rm int} U_0^{t} =
    U_0^{\dag t}
    e^{-i\chi H_{\rm int}} U_0^{t} =  e^{-i\chi H_{I}(t)} ,
  \end{aligned}
\end{align}
where we used Equation~\eqref{eq:14}
and $H_{I}(t) \coloneqq  U_0^{\dag t} H_{\rm int} U_0^{t} $
is the interaction Hamiltonian in the interaction picture.
If we solve Eq.~\eqref{eq:16}
for an arbitrary time we obtain
Dyson's formula for discrete time dynamics:
\begin{align}
\label{eq:evolutionconplete}
      & \ket{\psi(t')}_I = U_I(t',t)\ket{\psi(t)}_I, \\
      & U_I(t',t) \deq \ST \Big[\prod_{s=t}^{t'-1} U_{I}(s) 
      \Big] =
\ST \Big[
  \exp
\Big  (
    -i \chi \sum_{s=t}^{t'-1} H_I(t)
  \Big)
\Big],
\nonumber
\end{align}
where $\ST$ is the time ordering operator, such that $\ST[A(t_1)B(t_2)]=\theta(t_1-t_2)A(t_1)B(t_2)+\theta(t_2-t_1)B(t_2)A(t_1)$, $\theta(x)$ denoting the Heaviside function.
If it exists, the scattering operator in the interaction picture is
given by
\begin{align}
  \label{eq:17}
  S = \slim_{t \to +\infty } U_I(t,-t)
\end{align}
Equation~\eqref{eq:evolutionconplete} allows us to compute matrix
elements of the scattering operator
as a formal power series in the coupling constant $\chi$.
For a theory on a lattice with local
interactions $ H_{\rm int} \deq \sum_{x} H_{\rm int}(x)$
we have:
\begin{align} \label{eq:powerexpSmatrix}
  \bra{\psi}S\ket{\phi} =
    \sum_{n=0}^{+\infty} \frac{(-i\chi)^n}{n!}
    \sum_{t_j,x_j}
    \bra{\psi}  \ST[ \prod_{j=1}^n H_{I}(t_j,x_j)] \ket{\phi},
\end{align}
We observe that each term of the expansion
conserve the energy modulo $2\pi$
and, if the interaction is local, the total momentum is
conserved  modulo $2\pi$.
In many cases, $H_{\rm int}(x) $ is a polynomial
in the field operators and, from
Eq.~\eqref{eq:powerexpSmatrix}, we need to compute
time-ordered products of field operators.
The evaluation of
the terms appearing in the perturbation expansion of the
$S$-matrix can be performed by applying
Wick's theorem and can diagrammatically
be represented in terms of Feynman diagrams.
\begin{figure}[t]
  \centering
  \includegraphics[width=0.35\textwidth]{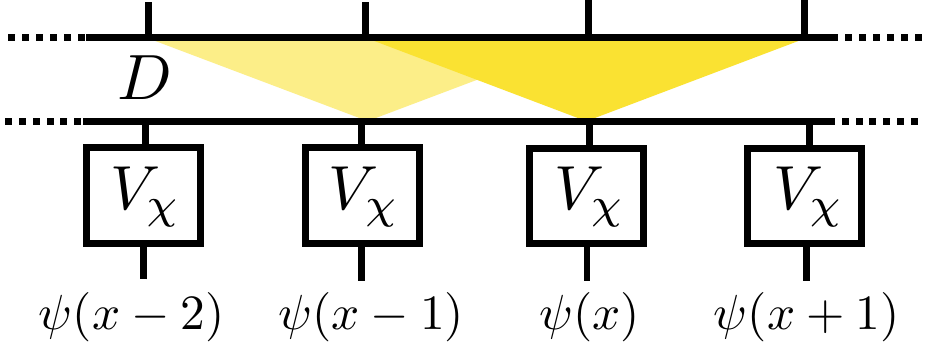}
  \caption{Thirring QCA unitary step. Each site of the
    lattice corresponds to a two-component fermionic field
    $\psi(x)$. The interaction $V_\chi$ is  completely local while
    the free evolution $D$ involves the nearest neighbors.}
    \label{fig:thirring}
  \end{figure}

\emph{The Thirring cellular automaton.}  We will
now apply the techniques of the previous
paragraphs to a one-dimensional Fermionic cellular automaton
with a four-Fermion on-site interaction, called Thirring quantum
cellular automaton~\cite{PhysRevA.97.032132}.
A two-component fermionic field $\psi \coloneqq (\psi_{\uparrow},
\psi_{\downarrow})^T$ is defined at every lattice site $x \in
\mathbb{Z}$ (see Fig. \ref{fig:thirring})
and its single step evolution is given by the unitary
operator:
\begin{align}
  \label{eq:6}
  \begin{aligned}
& U \coloneqq DV_{\chi}, \;  V_{\chi} \coloneqq e^{ i\chi
       \sum_{x\in \mathbb{Z}}\psi^{\dag}_{\uparrow}(x)
  \psi_{\uparrow}(x) \psi^{\dag}_{\downarrow}(x)
  \psi_{\downarrow}(x)  },\\
&D^{\dagger} \psi_{\uparrow} (x,t-1) D = \nu
\psi_{\uparrow} (x - 1,t) -i \mu
\psi_{\downarrow} (x,t),\\
&D^{\dagger} \psi_{\downarrow} (x,t-1) D = \nu
\psi_{\downarrow} (x + 1,t) -i \mu \psi_{\uparrow} (x,t),\\
  & \nu,\mu \in [0,1],\;\; \nu^2+\mu^2=1, \;\; \chi \in
  (-\pi,\pi ].
  \end{aligned}
\end{align}
The vacuum state of the model and a basis for the Fock
space are defined as follows:
\begin{align}
  \label{eq:25}
  \begin{aligned}
&     \ket{\Omega} \mbox{ s.t. } \psi_{\uparrow} (x)
  \ket{\Omega} = \psi_{\downarrow} (x)
  \ket{\Omega} = 0 \;\; \forall\,\,  x \in \mathbb{Z},\\
  &\ket{{a_1},{x_1}; \dots; a_n,x_n } \coloneqq \psi_{a_1}^{\dag}(x_1)\ldots \psi_{a_n}^{\dag}(x_n) \ket{\Omega}.
  \end{aligned}
\end{align}

The unitary operator $D$
describes the free
evolution  (occurring in discrete time steps) of
massive Dirac fermions on a one dimensional lattice and can
be diagonalised as:
\begin{align}
  \label{eq:24}
  \begin{aligned}
    &    D = e^{-i H_D}, \;
    H_D \coloneqq \int_{-\pi}^{\pi} \!\!\! dk
     \sum_{s=\pm} s \omega(k)
    \psi^{\dagger}_s(k) \psi_s(k) \\
 & \psi_s(k)\coloneqq\sum_{x \in \mathbb{Z}} \frac{e^{-ikx}(\mu \psi_{\uparrow} (x)  + g_s(k) \psi_{\downarrow} (x) )}
  {(2\pi)^{1/2}|N_s(k)|}
  \end{aligned}
\end{align}
where $\omega(k) \coloneqq \arccos(\nu \cos k)$,
$|N_s(k)|^2 = \mu^2 + |g_s(k)|^2$ and
$g_s(k)= s \sin \omega(k) + \nu \sin k$.  The non linear
evolution $ V_{\chi}$ is characterized by the four-fermion
interaction of the Thirring and the Hubbard models~\cite{THIRRING195891,Hubbard238,PhysRevD.11.2088,essler2005one}. We
notice that, as a consequence of definition~\eqref{eq:25},
the Hamiltonian $ H_D$ is not positive definite (hence
the states of Equation~\eqref{eq:25} are sometimes referred
to as \emph{pseudoparticle} states). As the full
evolution $U$ preserves the number of particles, it is
convenient to study the dynamics in this representation.

The two particle sector of this quantum cellular automaton
can be analytically solved~\cite{PhysRevA.97.032132} and it
is an ideal test for the perturbative methods
previously introduced.
We can show \cite{supplement} that  the matrix elements of $S$ reads as follows:
\begin{multline} \label{eq:28}
  \bra{k'_1,s'_1; k'_2,s'_2} S-I \ket{k_1,s_1; k_2,s_2} = \\
    \delta_{4\pi}(2p-2p')
    \delta_{2\pi}(\omega - \omega')
    \sum_{n=0}^{\infty} (e^{i \chi} -1)^{n+1} \gamma_n,
\end{multline}
where we defined
$\omega \coloneqq s_1\omega(k_1) +s_2\omega(k_2)  $,
$\omega' \coloneqq s'_1\omega(k'_1) +s'_2\omega(k'_2)  $
and  $\gamma_n$ are suitable coefficient which
depends on $k_i , s_i, k'_i,s'_i$.
From Equation~\eqref{eq:28} it is clear that 
processes in which there is a transition between
different values of the quasi-energy are allowed.
The scattering processes for the Thirring automaton can also
be perturbatively evaluated by applying Equation~\eqref{eq:powerexpSmatrix}.
The terms of the perturbative expansion can be labeled by
Feynman diagrams. We have e.g.
\begin{align} \nonumber
  \begin{aligned}
   &   \bra{(k'_1 +\pi), \,  - \, ; (k'_2 + \pi), \, - \,
   }S-I\ket{k_1, \, + \, ;
  k_2, \, + \,} = \\
   &= \vcenter{\hbox{
      \includegraphics[width=0.0325\textwidth]{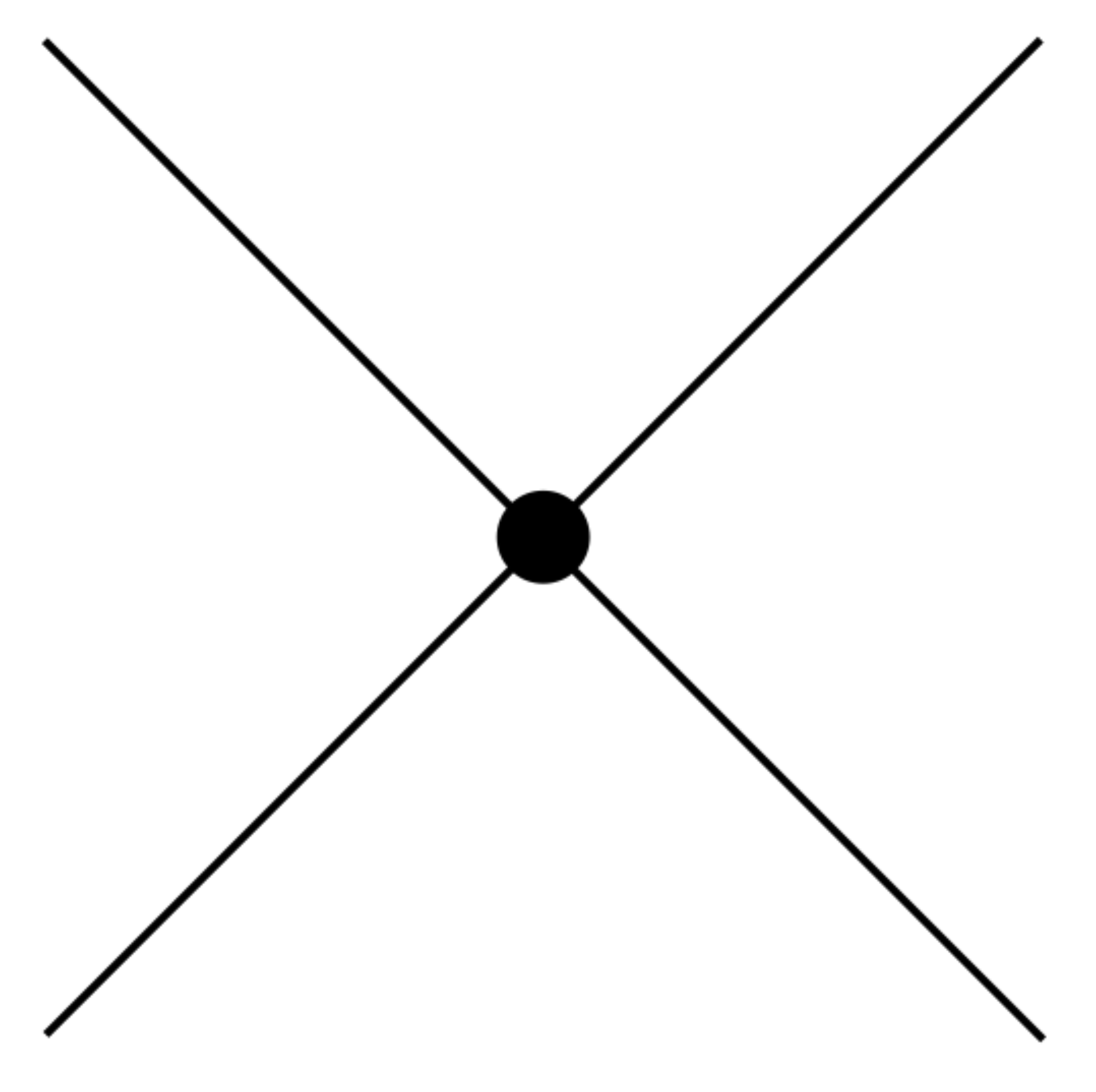}
    }}
    \: \chi +
    \vcenter{\hbox{
      \includegraphics[width=0.065\textwidth]{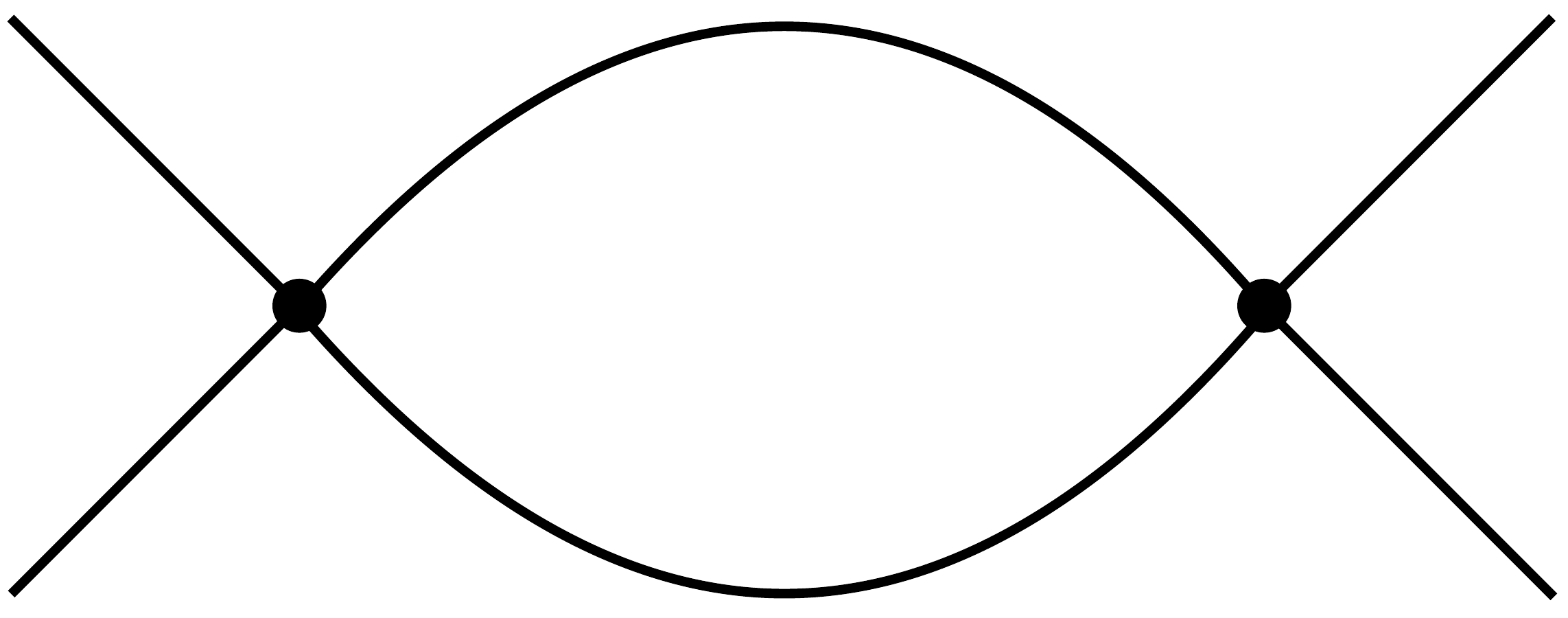}
    }}
  \: \chi^2 + \dots =\\
   &= \delta(k_1-k_1') \delta(k_2-k_2') ( i h_{k_1,k_2}    \, \chi 
+   2 h_{k_1,k_2}^2\, \chi^2 + o(\chi^2)) .
  \end{aligned}
\end{align}
where $h_{k_1,k_2}$ are suitable coefficients.
We notice the following technical detail which
has no counterpart in the continuos case: the iterative solution of the Lippmann-Schwinger equation
leads to a power series in the variable
$\lambda=e^{i\chi}-1$, while the Dyson series is an
expansion in the variable $\chi$. By expanding 
$\lambda$ in powers of $\chi$ one can check that the two
approaches agree.

\emph{Discussion.}  In this work we analyzed scattering
processes for quantum systems which evolve in discrete time
steps. We showed how to adapt some of the theoretical tools
of continuous-time systems to the discrete case. Both the
expansion of the Lippmann-Schwinger equation and the
Dyson
series maintain a formal analogy with their continuous
counterparts.  However, some intuitions must be modified,
most notably the notion of energy is replaced by
quasi-energy: just like momentum in a lattice is defined
only up to a reciprocal lattice vector, if time is discrete
energy is defined only modulo $2 \pi/ \tau$ ($\tau$ is the
time step).  This feature is also characteristic of
periodically driven quantum systems.  The periodicity in
energy conservation allows for a wider multiplicity of
scattering processes as we have seen in the case of the
Thirring automaton.

We also discussed how the scattering amplitudes of
the Trotterized model can recover the ones of a
continuous-time theory.  A rather exhaustive
assessment, conveyed by Equations \eqref{eq:9} and
\eqref{eq:expanddifferencescattering},
can be given if the free Hamiltonian
is bounded, as in a lattice theory with bounded
number of particles.  In this case, if the time
step $\tau$ is sufficiently small (see Equation
\eqref{eq:9}), the scattering between
different values of energy are suppressed and
the discrete-time scattering amplitudes equal
the amplitudes of the continuous
model plus an analytic function of $\tau$ which
vanishes in the limit $\tau \to 0$.  In the case
of unbounded spectrum, the suppression of
scattering between different energy states cannot
be established from the outset but it requires a
deeper examination of the dynamics, a necessary
condition being a rapid decay of the Fourier
transform of the potential.  

The application of the present theoretical
framework to the simulation of scattering in a
relativistic quantum field theory raises technical
issues, mainly because these models are only
defined via a renormalized (asymptotic)
perturbative series of the scattering amplitudes.
The existing approaches (as
the one of Ref. \cite{jordan2012quantum}) first
consider the simulation (for a finite time $T$)
of a continuous-time lattice
Hamiltonian dynamics, then extrapolate a
continuum limit by considering smaller and smaller
lattice spacings. Following the ideas presented in
the present paper, 
one could $i)$ compare the scattering
operator of the continuous time lattice theory and
the scattering operator of the discrete time
theory, and then $ii)$ extrapolate a (renormalised)
continuous limit.

\begin{acknowledgments}
  We acknowledge the support of the John Templeton
  Foundation under the project ID\# 60609 Quantum Causal
  Structures. The opinions expressed in this publication are
  those of the authors and do not necessarily reflect the
  views of the John Templeton Foundation.
\end{acknowledgments}

\bibliographystyle{apsrev4-1}
\bibliography{bibliography}

\clearpage

  \section*{\Large{SUPPLEMENTAL MATERIAL}}

\section{Derivation of the Lippmann-Schwinger Equation}
In this section we prove Eq.  \eqref{eq:5} of the main text.
First let us write the operators $\Omega_\pm$ as follows:
\begin{align}
  \label{eq:S2}
  \begin{aligned}
  \Omega_- = \lim_{n \to + \infty} \sum_{j=0}^{n-1}
  (U^{\dag j+1}U_0^{j+1} -U^{\dag j}U_0^{j} ) + I= \\
  = \lim_{n \to + \infty} \sum_{j=0}^{n-1}
  U^{\dag j}(U^\dag U_0-I) U_0^{j} ) + I
\end{aligned} \\
  \label{eq:Somega+}
  \begin{aligned}
     \Omega_+ = \lim_{n \to + \infty} \sum_{j=1}^{n}
  (U^{j}U_0^{\dag j} -U^{j-1}U_0^{\dag j-1} ) + I= \\
  = -\lim_{n \to + \infty} \sum_{j=-n}^{-1}
  U^{\dag j}(U^\dag U_0-I) U_0^{j}  + I.
  \end{aligned}
\end{align}
By  inserting the spectral resolution of $U_0$ 
into Equation
\eqref{eq:Somega+} we obtain
\begin{align}
  &\Omega_+ -I = \lim_{n \to + \infty} \sum_{j=0}^{n-1}
    U^{j}(U- U_0) U_0^{\dag j+1}=
  \nonumber \\
&  \lim_{\epsilon \to 0^+} \int_{B} \!\!\! dk  \sum_{j=0}^{\infty}
 e^{(i \omega(k) -\epsilon)(j+1)} U^j (U-U_0 )
 \ketbra{k}{k}  = \nonumber \\
 &\lim_{\epsilon \to 0^+} \int_{B} \!\!\! dk G(e^{-i
   \omega(k) +\epsilon}) U_0 (U_0^\dag U-I )\ketbra{k}{k},
   \label{eq:Somegalippman}
\end{align}
where we inserted a convergence factor.

From Eq.  \eqref{eq:3} and reminding that
$\Omega_{\pm}$ are
isometric we have
\begin{align}
  \label{eq:S1}
  S- I = \Omega_{-}^\dag \Omega_{+} -
  \Omega_{+}^\dag\Omega_{+} =
  (\Omega_{-}- \Omega_{+})^\dag\Omega_{+} .
\end{align}
By inserting Equation~\eqref{eq:S2} and
Equation~\eqref{eq:Somega+} into Equation \eqref{eq:S1} we
obtain
\begin{align}
  \label{eq:S expanded}
  \begin{aligned}
     S- I &= \lim_{n \to \infty} \sum_{j=-n}^{n-1}
     U_{0}^{\dag j} (U_{0}^{\dag }U-I)U^{j} \Omega_+ = \\
    &=  \lim_{n \to \infty} \sum_{j=-n}^{n-1}
     U_{0}^{\dag j} (U_{0}^{\dag }U-I) \Omega_+ U_0^{j},
  \end{aligned}
\end{align}
where in the last equality we used the intertwining relation
  \eqref{eq:2}.
  Then, by inserting Equation \eqref{eq:Somegalippman}
  into Equation \eqref{eq:S expanded}
  we have:
  \begin{align}
    \begin{aligned}
&        \bra{k'}S-I\ket{k} =
    \sum_{j=-\infty}^{+\infty} \bra{k'}
     U_{0}^{\dag j} (U_{0}^{\dag }U-I) \Omega_+
    U_0^{j}\ket{k} =\\
  &  \sum_{j=-\infty}^{+\infty} e^{i (\omega(k')-\omega(k) )j}\bra{k'}
    (U_{0}^{\dag }U-I) \Omega_+ \ket{k} =\\
    &2\pi \,\delta_{2 \pi }(\omega(k')-\omega(k))
   \lim_{\epsilon \to 0^+} \bra{k'}  T(e^{-i
   \omega(k) +\epsilon})\ket{k},
    \end{aligned}
  \end{align}
  where $G(z)= (zI-U)^{-1}$ is the resolvent operator of $U$
  and we defined
  \begin{align}
    \label{eq:SdefT}
    \begin{aligned}
      T(z) &:= W+ W G(z) U_0
    W,\\
   W &:=  U_0^\dag U-I.
    \end{aligned}
  \end{align}

  \section{Discrete-time vs continuous-time scattering}

We consider the simplest case of
scattering of a  single particle against a
potential $V$. 
The discrete time evolution is given by:
\begin{align*}
  U_0 &:= e^{-iH_0 \tau} \quad   U := e^{-iH_0 \tau}e^{-iV\tau}
\end{align*}
where $H_0$ has spectral resolution
\begin{align*}
   H_0 &:=
         \int_{0}^{{\omega_M}}
         \!\! \!\!
         \omega \, 
    dP_\omega ,
\end{align*}
$V$ is bounded 
and $\tau$ is the size of the temporal
step.
The corresponding scattering operator $S^{(\tau)}$can
be written as follows
\begin{align*}
   \bra{k'}S^{(\tau)}-I\ket{k} =& -2 \pi i\,
     \sum_{l\in \mathbb{Z}} \delta\Big(
                         \omega_{k'}-\omega_k
                         +2\pi \frac{l}{\tau}\Big) \\
\times &\bra{k'} \widetilde{T}^{(\tau)}(e^{-i(\omega_k + i\epsilon)\tau})
  \ket{k},
 \end{align*}
where $\widetilde{T}^{(\tau)}(e^{-i(\omega_k +
  i\epsilon)\tau})$ obeys
\begin{align}
  &\begin{aligned}
  \widetilde{T}^{(\tau)}
  (e^{-i(\omega_k +i\epsilon)\tau}) =
  \widetilde{W}^{(\tau)} + &\\
  +\widetilde{W}^{(\tau)}
  \widetilde{G}_0^{(\tau)}
  (e^{-i(\omega_k +i\epsilon)§\tau})&
\widetilde{T}^{(\tau)}
                                    (e^{-i(\omega_k +i\epsilon)\tau}),    
  \end{aligned}
\label{eq:SLIPSCHforTTilde}  \\
  &\begin{aligned}
  \widetilde{W}^{(\tau)}&:= \frac{i}{\tau}\Big(
  e^{-iV \tau}-I\Big),
  \end{aligned}
  \\
  &\begin{aligned}
  \widetilde{G}_0^{(\tau)}
  (e^{-iz\tau})&:=
  -i\tau \Big(  e^{-iz\tau} -
                  e^{-iH_0 \tau} \Big)^{-1}
                  e^{-iH_0 \tau}=\\
  &=-i\tau \int_{0}^{{\omega_M}}
         \!\! \!\!
  \frac{e^{-i\omega \tau} }
  {e^{-i z \tau}-e^{-i\omega \tau}}\, 
  dP_\omega  =\\
  &= \int_{0}^{{\omega_M}}
         \!\! \!\!
  \frac{-i\tau }
  {e^{-i(z - \omega)\tau}-1}\, 
  dP_\omega     \label{eq:greendiscretespectres} 
  \end{aligned}
\end{align}
where the limit $\epsilon \to 0^+$ has been left implicit.
On the other hand, the scattering operator $S$
for continuous time model with Hamiltonian $H = H_0
+V$ is given by
\begin{align}
  \bra{k'}S-I\ket{k} &= -2 \pi i\,
     \delta(
                         \omega_{k'}-\omega_k)
                        \bra{k'} T(\omega_k + i \epsilon)
         \ket{k}, \nonumber \\
T(\omega_k + i \epsilon) &=
  V + V G_0(\omega_k + i \epsilon)
  T(\omega_k + i \epsilon) \label{eq:LIPSCHforTinthecontinu} \\
  G_0(\omega_k + i \epsilon) &:=
                               (\omega_k + i
                               \epsilon -
                               H_0)^{-1} =
                               \nonumber \\
  &=\int_{0}^{{\omega_M}}
         \!\! \!\!
\frac{1}{\omega_k + i \epsilon -\omega}  \, 
  dP_\omega,\nonumber
\end{align}

We now make the following assumptions:
\begin{enumerate}
  \item $ 
    \omega_M < +\infty$ which holds
    for any nearest-neighbour,
    translational-invariant  free Hamiltonian of a
    theory on a lattice.
    \label{item:assumpt1}
\item $| V G_0(\omega_k + i \epsilon)|=\gamma <
  1$. This condition guarantees the convergence of
  the Born series.\label{item:assumpt2}
\end{enumerate}

Let us now evaluate the quantity
$\bra{k'}S -
S^{(\tau)}\ket{k}|$ as a
function of $\tau$.

Thanks to assumption \ref{item:assumpt1}
we have, for any $\delta > 0$ 
\begin{align}
 & \tau < \frac{2\pi}{\omega_M + \delta } \implies \nonumber \\
  &\implies \sum_{l\in \mathbb{Z}} \delta\Big(
                         \omega_{k'}-\omega_k
                         +2\pi \frac{l}{\tau}\Big)
  =
\delta(
                         \omega_{k'}-\omega_k
    ) \implies \nonumber\\
       & 
          \bra{k'}S_\tau-I\ket{k} = -2 \pi i\,
    \delta(
                         \omega_{k'}-\omega_k)
                       ) 
 \bra{k'} \widetilde{T}^{(\tau)}(e^{-i(\omega_k + i\epsilon)\tau})
  \ket{k},\nonumber
\end{align}

Let us now write
Equation~\eqref{eq:greendiscretespectres}
as follows
\begin{align*}
&\int_{0}^{{\omega_M}}
         \!\! \!\!
  \frac{-i\tau }
  {e^{-i(z - \omega)\tau}-1}\, 
  dP_\omega  =\\
  &=\int_{0}^{{\omega_M}}
         \!\! \!\!
\frac{1}{z-\omega}  E_{B_n}(-i(z-\omega)\tau )\, 
  dP_\omega,
\end{align*}
where $ E_{B_n}(x)$ is the exponential generating
function of the Bernoulli numbers $B_n$ :
\begin{align}
  \label{eq:bernoulli}
  E_{B_n}(x) = \sum_{n=0}^{\infty}
  \frac{x^n}{n!}B_n, \quad |x| < 2\pi
\end{align}
Therefore we have
\begin{align*}
&  \frac{1}{z-\omega}  E_{B_n}(-i(z-\omega)\tau
  )=
  \frac{1}{z-\omega}
                 -\frac{i\tau}{2}-\frac{\tau}{2}
                 f\left(\frac{(z-\omega)\tau}{2}\right)\\
  &f(x) :=\frac{1}{x}\Big( 1-\frac{x}{\tan(x)}
    \Big) \mbox{ for }0<|x|<\pi, \quad f(0):=0,
\end{align*}
where $ f(x)$ is holomorphic for
$|x| < \pi$.
Let us fix an arbitrary $0<\delta' < \delta$. Then, in the compact set
$(\tau,\omega,\epsilon) \in [0, \frac{2\pi}{\omega_M + \delta
}]\times [0, \omega_M]\times
[0,\delta']$, the function
$
f\left(\frac{(\omega_k-\omega+i\epsilon)\tau}{2}\right)$
is smooth and bounded and it
converges uniformly to
$f\left(\frac{(\omega_k-\omega)\tau}{2}\right)$.
We can then write
\begin{align}
&  \widetilde{G}_0^{(\tau)}(e^{-i(\omega_k
  +\epsilon)\tau}) = G_0(\omega_k+i \epsilon)
                -\frac{i\tau}{2}-\frac{\tau}{2}F(\omega_k,\tau)
  \label{eq:greendisexp2}\\
  &F(\omega_k,\tau):=\int_{0}^{{\omega_M}}
         \!\! \!\!
    f\left(\frac{(\omega_k-\omega)\tau}{2}\right)\,
    dP_\omega \nonumber
\end{align}
and we have
\begin{align}
  |F(\omega_k,\tau)|   &= \max_{ \omega } 
                         \left|f\left(\frac{(\omega_k-\omega)\tau}{2}\right)\right|
                         \leq   \nonumber \\
                       &\leq
                         f\left(\frac{\omega_M
                         \tau}{2}\right),
                         \label{eq:boundforf}
\end{align}
where the last inequality follows from the fact
that $f(x)$ is odd and non decreasing.
Then, for
$\tau \in [0, \frac{2\pi}{\omega_M + \delta}]$,
$ F(\omega_k,\tau)$ is a bounded operator
 and it is 
analytic in $\tau$.

Let us now write $ \widetilde{W}^{(\tau)}$
as follows,
\begin{align}
  \widetilde{W}^{(\tau)} = V +\tau\, Q(\tau)\, V^2 \label{eq:potentialexpanded}\\
  Q(\tau) :=
  -i\sum_{n=0}^{\infty}\frac{(-iV\tau)^n}{(n+2)!} \nonumber
\end{align}
where $Q(\tau)$ is bounded an analytic for any $\tau$.
For example, if $V$ is a local potential, we
have:
\begin{align*}
  V &= \int_{\supp(v)} \!\!\!\!\!\!
  \!\!\!\!\!\!\!\!\! dx\, v(x) \ketbra{x}{x}
  \implies\\
  Q(\tau) &= \int_{\supp(v)} \!\!\!\!\!\!
  \!\!\!\!\!\!\!\!\! dx\, q(\tau v(x)) \ketbra{x}{x} \\
  q(y) &:=
  \frac{-i - y + ie^{-i y }}
  { y^2},
\end{align*}
where $\supp(v)$ denotes the support of the
real-valued function $v$, whose domain can be either
$\mathbb{R}^n$ or a discrete lattice (in this
latter case  the integral should be replaced by a
discrete sum).
Since $|q(y)|^2$ is an even function and it is
decreasing in $y \in [0 , +\infty)$ 
we have that
\begin{align}
  |Q(\tau)| = \sup_{\tau,x} |q(v(x))| \leq |q(0)|
  = \frac12
  \label{Seq:boundforQ}
\end{align}
Combining
Equation~\eqref{eq:greendiscretespectres} and
Equation~\eqref{eq:potentialexpanded} we obtain:
\begin{align}
 \widetilde{W}^{(\tau)}&  \widetilde{G}_0^{(\tau)}
                           =\nonumber
  \\
  &\begin{aligned}
  =&\Big(   V +\tau\, Q  \,V^2\Big)\Big( G_0
  -\frac{i\tau}{2}-\frac{\tau}{2}F\Big)
= 
\end{aligned}\nonumber \\
   &\begin{aligned}
        =&VG_0 +
        \tau R_1 + \tau^2 R_2,  \end{aligned}\label{eq:tldeQtimestildeWexpand}\\
      &\begin{aligned}
        R_1 &:= 
        \frac{i}{2}V - \frac{1}{2} VF+
        Q\, V^2 \,G_0 \\
         R_2&:=
        -\frac{i}{2}\, Q \,V^2
         - \frac{1}{2}  Q\, V^2\, F ,
  \end{aligned}\label{eq:tldeQtimestildeWexpand2}
\end{align}
where we use the shorthand notation
$ \widetilde{G}_0^{(\tau)} := \widetilde{G}_0^{(\tau)}(e^{-i(\omega_k +
  i\epsilon)\tau})$ and
$ {G}_0 := {G}_0(\omega_k +
  i\epsilon)$
From assumption \ref{item:assumpt2} we have that
$|VG_0|=\gamma<1 $.
By a straightforward application of the triangle
inequality in Equation~\eqref{eq:tldeQtimestildeWexpand}
we have that
\begin{align}
|  \widetilde{G}_0^{(\tau)}
                            \widetilde{W}^{(\tau)}|
  &\leq
  \gamma + \tau\gamma' + \tau^2
    \gamma'' \label{eq:SboundforGW3}\\
  \gamma'& := \frac{1}{2}|V|(1 + |F|+
  \gamma |Q|) \nonumber \\
\gamma'' &:= \frac{1}{2} |V|^2\, |Q|(1+ |F|) \nonumber
\end{align}
We have then proved that
$ \widetilde{W}^{(\tau)}\widetilde{G}_0^{(\tau)}$
is a bounded operator valued analytic function.
Let us now estimate which range of value of $\tau$
guarantees that
$|\widetilde{W}^{(\tau)}\widetilde{G}_0^{(\tau)} |<1$.
By inserting Equation \eqref{Seq:boundforQ} and
Equation~\eqref{eq:boundforf} into Equation
\eqref{eq:SboundforGW3} we have
\begin{align}
  | \widetilde{W}^{(\tau)} \widetilde{G}_0^{(\tau)}
                            |
  \leq \,&  \gamma +
  \tau
  \frac12 |V| \Big(1 +  f\left(\frac{\omega_M
                         \tau}{2}\right) +
  \frac{\gamma}{2} \Big)+ \nonumber \\
       &+\tau^2 \frac{|V|^2}{4}
         \Big(1+
         f\left(\frac{\omega_M\tau}{2}\right)\Big)
          \label{eq:boundGWsimplified}
\end{align}
Moreover, since
\begin{align*}
  f(x) \leq \frac{2}{\pi^2 x} \mbox{ for }x \in [0,\tfrac{\pi}{2}),
\end{align*}
we have that
\begin{align}
 |\widetilde{W}^{(\tau)}  \widetilde{G}_0^{(\tau)}
                            |
  \leq\, &\gamma +
  \tau \frac12 |V| \Big(1 +  f\left(\frac{\omega_M
                         \tau}{2}\right) +
  \frac{\gamma}{2} \Big)+ \nonumber\\
       &+\tau^2 \frac{|V|^2}{4}
         \Big(1+
         f\left(\frac{\omega_M\tau}{2}\right)\Big)
         \leq  \nonumber \\
   \leq\,      &\gamma +
  a_1\tau |V|
          +a_2\tau^2|V|^2 ,
          \label{eq:semifinalboundforGW}\\
            &a_1:=
           \frac{3\pi+4}{4\pi}\quad a_2:=\frac{\pi+2}{4\pi} \nonumber
\end{align}
where we also used $\gamma <1$.
After a rather straightforward
computation\footnote{One can easily verify that
  $\tfrac{a_1}{2a_2}(\sqrt{1+(1-\gamma)\tfrac{4a_2}{a_1}}-1)
> \sqrt{2-\gamma}-1$} we have
\begin{align}
  \begin{aligned}
  &\tau \leq m^*
  \implies|  \widetilde{G}_0^{(\tau)}
                            \widetilde{W}^{(\tau)}|<1
                            \\
                            &m^*:=\min
  \Big(\frac{\sqrt{2-\gamma}-1}{|V|} ,
  \frac{\pi}{\omega_M} \Big)  
  \end{aligned}\label{eq:SfinalboundforGW}
\end{align}
From assumption \ref{item:assumpt2} and Equation
\eqref{eq:SfinalboundforGW} we can solve
Equation~\eqref{eq:SLIPSCHforTTilde} and
Equation
\eqref{eq:LIPSCHforTinthecontinu}                            
as follows:
\begin{align}
  \label{eq:SS2}
  \begin{aligned}
  \widetilde{T}^{(\tau)}
  =&
  (I-\widetilde{W}^{(\tau)}
  \widetilde{G}_0^{(\tau)}
  )^{-1}\widetilde{W}^{(\tau)}\quad \mbox{for }\tau \leq m^*\\
    {T}
  =&
  (I-{V}
 {G}_0
)^{-1}V   
  \end{aligned}
\end{align}
where both
$(I-\widetilde{W}^{(\tau)}
  \widetilde{G}_0^{(\tau)}
  )^{-1}$ and $ (I-{V}
 {G}_0
)^{-1}$ are bounded operators and
the shorthand notation
$ \widetilde{G}_0^{(\tau)} := \widetilde{G}_0^{(\tau)}(e^{-i(\omega_k +
  i\epsilon)\tau})$ ,
$ {G}_0 := {G}_0(\omega_k +
i\epsilon)$,
$ \widetilde{T}^{(\tau)} := \widetilde{T}^{(\tau)}(e^{-i(\omega_k +
  i\epsilon)\tau})$ and
$ T := T(\omega_k +
i\epsilon)$
is understood.
Then we have
\begin{align}
  \widetilde{T}&^{(\tau)}-T = \nonumber
  \\
&  \begin{aligned}
= & (I-\widetilde{W}^{(\tau)}
  \widetilde{G}_0^{(\tau)}
  )^{-1}\widetilde{W}^{(\tau)} +\\
  &- (I-{V}
 {G}_0
  )^{-1}V   =    
  \end{aligned}\nonumber
  \\
 & \begin{aligned}
=&(I-\widetilde{W}^{(\tau)}
  \widetilde{G}_0^{(\tau)}
  )^{-1}(V +\tau\, V^2\, Q(\tau) ) + \\
  &- (I-{V}
 {G}_0
  )^{-1}V =    
  \end{aligned}\nonumber
  \\
 & \begin{aligned}
    =&\Big( (I-\widetilde{W}^{(\tau)}
  \widetilde{G}_0^{(\tau)}
  )^{-1} -  (I-{V}
 {G}_0
  )^{-1} \Big) V +\\
  &\tau\, (I-\widetilde{W}^{(\tau)}
  \widetilde{G}_0^{(\tau)}
  )^{-1}  V^2\, Q(\tau)  =
\end{aligned}\nonumber\\
     & \begin{aligned}
    =& (I-\widetilde{W}^{(\tau)}
  \widetilde{G}_0^{(\tau)}
  )^{-1} (\tau R_1 + \tau^2 R_2)
 (I-{V}
 {G}_0
  )^{-1} V +\\
  &+\tau\, (I-\widetilde{W}^{(\tau)}
  \widetilde{G}_0^{(\tau)}
  )^{-1}  V^2\, Q(\tau)  =
\end{aligned}\nonumber \\
& \begin{aligned}
    =& (I-\widetilde{W}^{(\tau)}
  \widetilde{G}_0^{(\tau)}
  )^{-1} (\tau R_1 + \tau^2 R_2)
 T +\\
  &+ \tau\, (I-\widetilde{W}^{(\tau)}
  \widetilde{G}_0^{(\tau)}
  )^{-1}  V^2\, Q(\tau)  =
\end{aligned}   \nonumber\\
 &= \tau \,\widetilde{R}^{(\tau)} ,\\
               & \begin{aligned}
                     &\widetilde{R}^{(\tau)} := 
    (I-\widetilde{W}^{(\tau)}
  \widetilde{G}_0^{(\tau)}
  )^{-1} (R_1 + \tau R_2)
 T +\\
  &+ (I-\widetilde{W}^{(\tau)}
  \widetilde{G}_0^{(\tau)}
  )^{-1}  V^2\, Q(\tau)  
\end{aligned} \nonumber
\end{align}
where $\widetilde{R}^{(\tau)}$ is a bounded
operator.
Since $\widetilde{T}^{(\tau)}$ is also analytic in
$\tau$
we can compute the following first order
approximation:
\begin{align}
  \label{eq:firstorder}
  \widetilde{T}^{(\tau)}-T = -i\tau\,  T \,V + O(\tau^2)
\end{align}
which gives
\begin{align}
  \begin{aligned}
     \bra{k'}S^{(\tau)}-S\ket{k}  =&- 2 \pi \tau \,
    \delta(
                         \omega_{k'}-\omega_k
                         ) \cdot \\
  &                       
\cdot \bra{k'} {T}(\omega_k + i\epsilon)V
  \ket{k}+O(\tau^2).
  \end{aligned}
\end{align}

\section{Thirring Quantum Cellular Automaton}

The Thirring QCA is a
one-dimensional massive
Fermionic cellular automaton with a four-Fermion on-site
interaction.
At any lattice point $x\in\mathbb{Z}$
we have a two-component
Fermionic field $\psi$ defined as follows:
\begin{align}
  \begin{aligned}
   & \psi(x)=
    \begin{pmatrix}
      \psi_\uparrow(x)\\
      \psi_\downarrow (x)\end{pmatrix},
    \\
 &   [ \psi_a(x),      \psi_b (y)]_+ =
    [ \psi^\dag_a(x),      \psi^\dag_b (y)]_+ = 0 ,\\
 &   [ \psi_a(x),      \psi^\dag_b (y)]_+ = \delta_{ab}
 \delta_{xy}, \quad a,b \in \{\uparrow, \downarrow \}.
  \end{aligned}
\end{align}
We then define the vacuum state and a basis for the Fock
space as follows:
\begin{align}
  \label{eq:S25}
  \begin{aligned}
&     \ket{\Omega} \mbox{ s.t. } \psi_{a} (x)
  \ket{\Omega} = 0, \; \;\; \forall\,\,  x \in \mathbb{Z},\;\;
  \forall \,\, a
  \in  \{\uparrow, \downarrow \},\\
  &\ket{{a_1},{x_1}; \dots ; a_n,x_n } := \psi_{a_1}^{\dag}(x_1)\cdots \psi_{a_n}^{\dag}(x_n) \ket{\Omega}.
  \end{aligned}
\end{align}
This choice is sometimes referred to as the \emph{pseudo
  particle} representation. The evolution has the form
\begin{align}
  \label{eq:SdefiningU}
U :=  DV_\chi ,
\end{align}
where $D$ is the free evolution and 
$V_\chi$ is the interaction term.
The free evolution $D = e^{-i H_D}$ is defined as
\begin{align}
  \label{eq:SdefiningD}
  &\begin{aligned}
  &D^{\dagger} \psi_{\uparrow} (x,t-1) D = \nu
\psi_{\uparrow} (x - 1,t) -i \mu
\psi_{\downarrow} (x,t),\\
&D^{\dagger} \psi_{\downarrow} (x,t-1) D = \nu
                             \psi_{\downarrow} (x + 1,t) -i
                             \mu \psi_{\uparrow} (x,t),   
  \end{aligned}
 \\
&\nu,\mu >0,\: \nu^2+\mu^2=1,
\end{align}
and $H_D$ can be diagonalised as follows:
\begin{align}
   \begin{aligned}
 & H_D := \int_{-\pi}^{\pi} \!\!\! dk
     \sum_{s=\pm} s \omega(k)
    \psi^{\dagger}_s(k) \psi_s(k), \\
    & \psi_\pm(k):=
\alpha_{\pm,\uparrow}(k) \psi_{\uparrow} (k) + \alpha_{\pm,\downarrow}(k) \psi_{\downarrow}  ,
\\
 & \psi_{a} (k) := \frac{1}{ \sqrt{2 \pi} }
  \sum_{x \in \mathbb{Z}}
  e^{-ikx} \psi_{a} (x),  \;\;\;
  a \in \{\uparrow, \downarrow   \}, 
  \\
  &\alpha_{s,\uparrow}(k) :=   \frac{\mu}{|N_s(k)|},
  \quad \alpha_{s, \downarrow}(k) :=     \frac{g_s(k)}{|N_s(k)|},\\
  &  \omega(k) := \arccos(\nu \cos k), \\
  & |N_s(k)|^2 = \mu^2 + |g_s(k)|^2, \\
  &g_s(k)= s \sin \omega(k) + \nu \sin k.
	\end{aligned} 
  \label{eq:SdiagonalizedD}
\end{align}
The interaction term $V_{\chi} $ is defined as follows:
\begin{align}
  \begin{aligned}
  &V_{\chi} := \exp  (i\chi H_{int}), \\
  & H_{int} :=     \sum_{x\in \mathbb{Z}}\psi^{\dag}_{\uparrow}(x)
  \psi_{\uparrow}(x) \psi^{\dag}_{\downarrow}(x)
  \psi_{\downarrow}(x),
  \end{aligned}
  \label{eq:SinteractioV}
\end{align}
where we notice the same on-site four-fermion interaction of
the Thirring’s and the Hubbard’s models.

We remark that both $D$ and $V_\chi$
commute with the total number operator
$n_\mathrm{tot}=\sum_{x\in\mathbb Z}(\psi_\uparrow^\dag(x) \psi_\uparrow
(x)+{\psi_\downarrow}^\dag (x) \psi_\downarrow(x))$.
Therefore, it is possible to analyze the dynamics for a
fixed number of particles.

\subsubsection{One-particle sector}

The
one-particle Hilbert space is
$\mathbb{C}^2\otimes \ell^{2}(\mathbb{Z})$
and we use the following
orthonormal basis:
\begin{align}
  \label{eq:S2parthilbert2}
  \begin{aligned}
 &    \ket{a}\ket{x} := \ket{ax} = \psi_a^{\dagger}(x)
  \ket{\Omega}, \;\;\; a \in \{ \uparrow,\downarrow \},\;\;
  x \in \mathbb{Z} \\
 & \ket{\uparrow} =
              \begin{pmatrix}
                1 \\
                0
              \end{pmatrix}, \;
  \ket{\downarrow} =
              \begin{pmatrix}
                0 \\
                1
              \end{pmatrix}.
  \end{aligned}
\end{align}
In this sector, the interaction term $V_\chi$
is irrelevant and the free evolution 
$D$ becomes a unitary operator $D^{(1)}$ on
$\mathbb{C}^2 \otimes
\ell^2(\mathbb{Z})$.
If we use the basis of Equation~\eqref{eq:S2parthilbert2}
we have
\begin{align}
  \label{eq:SdefiningD2}
  \begin{aligned}
    &  \ket{\psi(t+1)} = D^{(1)} \ket{\psi(t)} \\
    & D^{(1)}  :=
    \begin{pmatrix}
      \nu T^\dag&-i\mu\\
      -i\mu & \nu T
    \end{pmatrix},        
              \end{aligned}
\end{align}
where $T$ is the translation operator on
$\ell^2(\mathbb{Z})$, $T \ket{x} = \ket{x+1}$.
In the Fourier transformed basis we have:
\begin{align}
  \label{eq:SdefiningDfourier2}
  \begin{aligned}
    &   D^{(1)}  = \int_{-\pi}^{\pi} \!\!\! dk_1 \, D^{(1)}_{k_1}
\otimes
   \ketbra{{k_1}}{{k_1}} ,\\
&D^{(1)}_{k_1}:=
    \begin{pmatrix}
      \nu e^{ik_1}&-i\mu\\
      -i\mu & \nu e^{-ik_1}
    \end{pmatrix},        \\
  &  D^{(1)}_{k_1}\ket{u^{s}_{k_1}} = e^{-is \omega(k_1)}\ket{u^{s}_{k_1}},\\
 &   \ket{u^{s}_{k_1}} := \begin{pmatrix}
  \alpha_{s,\uparrow}(k_1)\\
  \alpha_{s,\downarrow}(k_1)
  \end{pmatrix},\\
  &\ket{k}:= \sum_{x \in \mathbb{Z}}
  e^{-ikx} \ket{x}.
              \end{aligned}
\end{align}
and one can verify that
$\ket{u^{s}_{k_1}} \ket{{k_1}}=
\psi_s^{\dag}({k_1})\ket{\Omega}$.  The unitary evolution
$D^{(1)}$ corresponds to the so-called one-dimensional Dirac
walk whose dynamics recovers that of a free Dirac field of
mass $\mu$ in the $k_1 \to 0$ limit.
\subsubsection{The two particle sector}
We now analyze the two particle sector of the Thirring QCA.
Since Thirring QCA is Fermionic model,
the Hilbert space $\mathcal{H}^{(2)}$ is the antisymmetric subspace of
$(\mathbb{C}^2\otimes
\ell^{2}(\mathbb{Z}))\otimes(\mathbb{C}^2\otimes
\ell^{2}(\mathbb{Z}))$, i.e.
\begin{align}
  \label{eq:SS1}
&  \mathcal{H}^{(2)} := \supp P_A,\quad P_A:=
  \frac{1}{2}(I - \Swap), \\
 & \Swap \ket{a_1}\ket{x_1}\ket{a_2}\ket{x_2} :=
  \ket{a_2}\ket{x_2}\ket{a_1}\ket{x_1} ,
\end{align}
and the evolution is given by
\begin{align}
  \label{eq:Sevol2particles}
  \begin{aligned}
    & U^{(2)} := P_A D^{(2)} V_{\chi}^{(2)}P_A
    \\
    & D^{(2)} :=   D^{(1)} \otimes D^{(1)} \\
    & V_{\chi}^{(2)} := \exp \left( i\chi \sum_{x \in
        \mathbb{Z}} I \otimes \ketbra{x}{x}\otimes I \otimes
      \ketbra{x}{x} \right).
  \end{aligned}
\end{align}
One can notice that both $D^{(2)} $ and
$ V_{\chi}^{(2)}$ commute with $P_A$. Therefore, it is
convenient to diagonalize the operator
$D^{(2)}V_{\chi}^{(2)}$ acting on
$(\mathbb{C}^2\otimes
\ell^{2}(\mathbb{Z}))\otimes(\mathbb{C}^2\otimes
\ell^{2}(\mathbb{Z}))$ and project its eigenfuctions on the
antisymmetric subspace afterwards.
Moreover, we notice that $D^{(2)} $, 
$ V_{\chi}^{(2)}$ and $P_A$
commute with the translation of the
centre of mass, i.e.
\begin{align}
  \label{eq:Scentrofmassispreserved}
  \begin{aligned}
     [  D^{(2)}  V_{\chi}^{(2)}, T_\mathrm{cm}] =0 \\
  T_\mathrm{cm}:=  I \otimes T
  \otimes I \otimes T
  \end{aligned}
\end{align}
It is then useful to introduce the centre of mass
coordinates
\begin{align}
  \label{eq:Scentrofmasscoordinates2}
  \begin{aligned}
      \ket{a_1} \ket{x_1} \ket{a_2}\ket{x_2}\to
  \ket{a_1,a_2}\ket{y}\ket{w}\\
 y = x_1 - x_2, \quad w = x_1 + x_2.
  \end{aligned}
\end{align}
We notice that only the pairs $y, w$ with $y$ and $w$ either
both even or odd, i.e. $(y,w) \in \mathsf{L}$ with
$ \mathsf{L} := \{ (y,w) \in \mathbb{Z}^2 | \in y+w = 2n, n \in
\mathbb{Z}\}$ correspond to lattice points in the original
basis $x_1, x_2$. However, it is convenient to let $y$ and $z$ 
run free and consider  another Hilbert space $\mathcal{H}'
\cong (\mathbb{C}^2\otimes
\ell^{2}(\mathbb{Z}))\otimes(\mathbb{C}^2\otimes
\ell^{2}(\mathbb{Z})) $.
The action of
$  D^{(2)} V_{\chi}^{(2)}$, $T_\mathrm{cm}$ and $P_A$  can be
straightforwardly extended to $\mathcal{H}'$
and the operator
\begin{align}
\nonumber
  P_{\mathsf{L}} : \ket{a_1,a_2}\ket{y}\ket{w} \mapsto
  \begin{cases}
  \ket{a_1} \ket{\tfrac{w+y}{2}} \ket{a_2}\ket{\tfrac{w-y}{2}}, &
  (y,w) \in \mathsf{L}\\
0    & \mbox{otherwise}
  \end{cases},
\end{align}
Projects on the Hilbert space of interest.
Correspondingly,
we introduce the Fourier transform basis
\begin{align}
  \label{eq:Scentrofmasscoordinatesfourier2}
  \begin{aligned}
  &\ket{y}\ket{w} \to \ket{k}\ket{p}, \quad p,k \in (-\pi,\pi]\\
 &\ket{k}:= \frac{1}{ \sqrt{2 \pi} }
  \sum_{y \in \mathbb{Z}}
  e^{-iky} \ket{y}, \\
&\ket{p}:= \frac{1}{ \sqrt{2 \pi} }
  \sum_{w \in \mathbb{Z}}
  e^{-ipw} \ket{w},   
  \end{aligned}
\end{align}
where 
$p$ and $k$ freely run in $(-\pi,\pi]$.
One can verify that the Fourier transformed operator $\tilde{P}_{\mathsf{L}}$ of
$P_{\mathsf{L}}$
acts on the improper states of the Fourier basis as follows:
\begin{align}
\nonumber
  \begin{aligned}
    \tilde{P}_{\mathsf{L}}&:
\ket{a_1,a_2}\ket{k}\ket{p} \mapsto
  \ket{a_1} \ket{k_1 } \ket{a_2}\ket{k_2},\\
 & k_1 := p+k \mod{2 \pi}, \quad k_2 := p-k \mod{2 \pi} 
  \end{aligned}
\end{align}
i.e.  the following diagram
\begin{align}
  \label{eq:S6}
  \begin{aligned}
  \includegraphics[width=7.3cm]{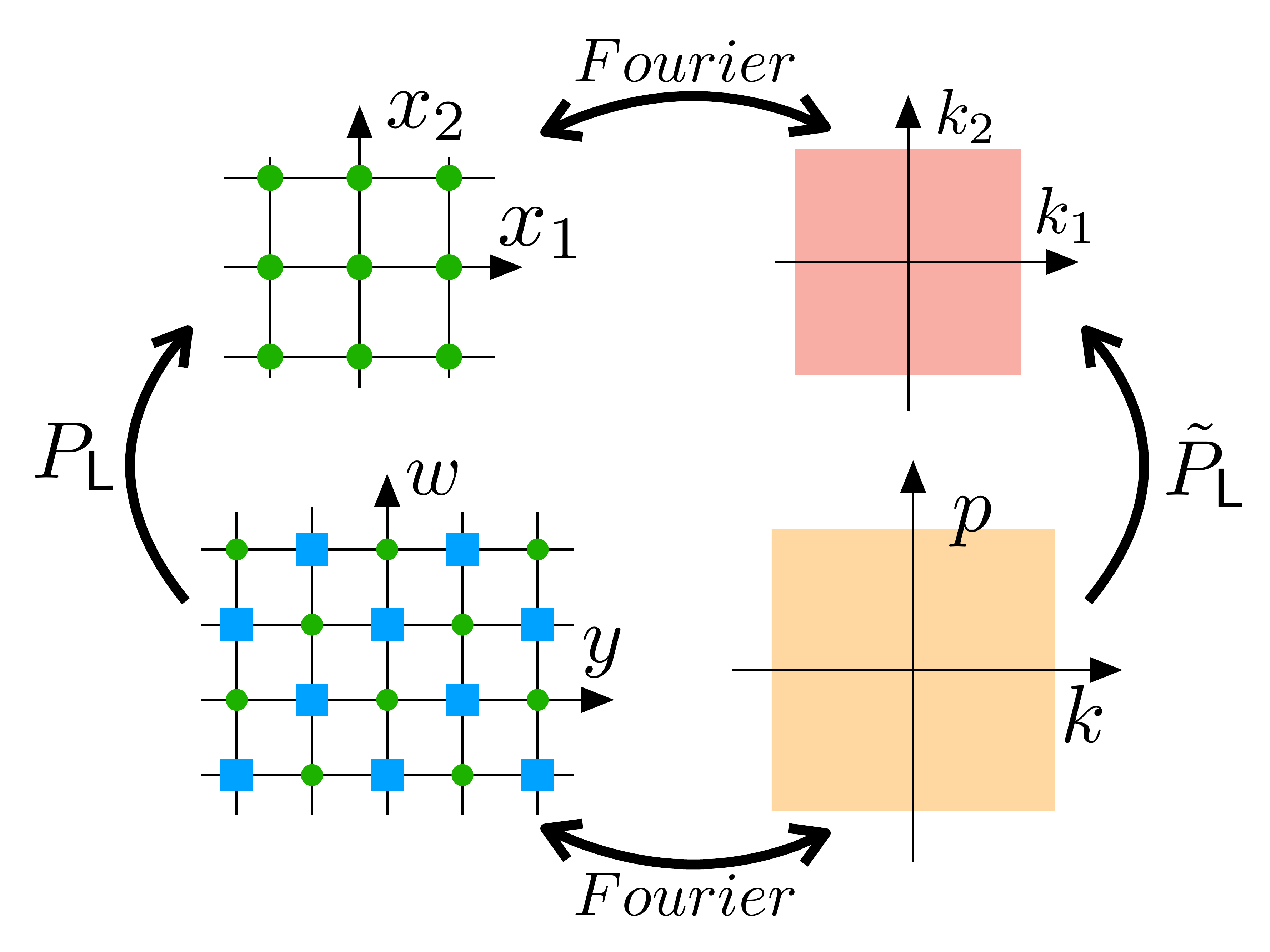}   
  \end{aligned}
\end{align}
commutes.

Then we have:
\begin{align}
  \label{eq:S20}
  \begin{aligned}
     &    D^{(2)} V^{(2)}  = \int_{-\pi}^{\pi} \!\!\!\! dp \,
     D^{(2)}_p \tilde{V}_\chi^{(2)}\otimes \ketbra{p}{p},\\
  &\tilde{V}^{(2)}_\chi := \sum_{y \in \mathbb{Z}}e^{i \chi
     \delta_{y,0}} I \otimes \ketbra{y}{y}, \\
  &  D^{(2)}_p = \int_{-\pi}^{\pi}  \!\!\!\!  dk  \, D_{p+k}^{(1)}\otimes
  D_{p-k}^{(1)} \otimes \ketbra{k}{k} ,\\
  &D^{(2)}_p \ket{v_{p,k}^{s_1s_2}}\ket{k} =  e^{-i \omega^{s_1,s_2}_{p,k}}
\ket{v_{p,k}^{s_1s_2}}\ket{k},\\
&\ket{v_{p,k}^{s_1s_2}} := \ket{u_{p+k}^{s_1}}\ket{u_{p-k}^{s_2} },\\
 &\omega^{s_1,s_2}_{p,k} := s_1\omega(p+k) + s_2\omega(p-k).
  \end{aligned}
\end{align}

In the centre of mass coordinates, the two particle
dynamics of the Thirring QCA reduces to the dynamics of a
single particle in the presence of a potential.
The Hilbert space of the system is $\mathbb{C}^4 \otimes
\ell^2(\mathbb{Z})$ and the evolution is given by the
following equation
\begin{align}
  \label{eq:Ssingleparticledelta}
  \ket{\phi(t+1)} = D^{(2)}_p \tilde{V}^{(2)}_\chi
  \ket{\phi(t)}.
\end{align}

\subsection{Scattering of two particles:
  Lippmann-Schwinger equation}

Let us now study the scattering operator for the unitary
evolution of Equation~\eqref{eq:Ssingleparticledelta}.
In this section we assume $p \neq n \frac{\pi}{2}$, $n \in
\mathbb{Z}$.
The analysis of the cases
$p = n \frac{\pi}{2}$,
which we omit,
can be carried out along the same lines.

As a preliminary step, let us compute the following
expression:
\begin{align}
  \label{eq:SWoperator}
  \begin{aligned}
     W := &{D^{(2)}_p}^\dag  {D^{(2)}_p} \tilde{V}^{(2)}_\chi
     -I =\\
     =&\tilde{V}^{(2)}_\chi -I =\lambda C
   \\
\lambda :=      & e^{i\chi}-1,  \quad C:= I \otimes
      \ketbra{0}{0}
  \end{aligned}
\end{align}

From Equation~\eqref{eq:12} in the main text, the formal
solution of the Lippmann-Schwinger equation for the $T$
matrix is the following:
\begin{align}
  \label{eq:STmstrix2}
  \begin{aligned}
      T(z) = &\sum_{n=0}^{\infty} (W (zI-D^{(2)}_p)^{-1}
  D^{(2)}_p)^nW =\\
  =&\sum_{n=0}^{\infty} \lambda^{n+1} ( C (z
  D^{(2)\dag}_p-I)^{-1} C)^n.
  \end{aligned}
\end{align}
By using Equation~\eqref{eq:S20} and Equation~\eqref{eq:SWoperator}
we obtain
\begin{align}
  \label{eq:Stherelevantmatrix2}
  &    C (z
  D^{(2)\dag}_p-I)^{-1} C 
  =
\Gamma(z) \otimes \ketbra{0}{0}, \\
&\Gamma(z) :=   \frac{1}{2 \pi} \int_{-\pi}^{\pi}  \!\!\!\!  dk  \, (zD_{p+k}^{(1)\dag}\otimes
    D_{p-k}^{(1)\dag} -I)^{-1}.\label{eq:Stherelevantmatrix3}
\end{align}
The improper matrix elements of the scattering matrix are
given by the following equation:
\begin{align}
  \label{eq:Sfullmatrixelements}
 & \begin{aligned}
 \bra{\Psi_{p,k}^{s_1s_2}}  &  S-I
\ket{\Psi_{p,k'}^{s'_1s'_2}}=\\
  & \lim_{\epsilon \to 0^+} \,
      \delta_{2 \pi}(\omega_{p,k}^{s_1s_2}-\omega_{p,k'}^{s'_1s'_2}) \cdot\\
  \cdot & 
          \sum_{n=0}^{\infty} \lambda^{n+1}\bra{w_{p,k}^{s_1s_2}}\Gamma^n(e^{-i \omega_{p,k}^{s_1s_2} +
          \epsilon})
          \ket{w_{p,k'}^{s'_1s'_2}}    
  \end{aligned}
  \\
&\ket{\Psi_{p,k}^{s_1s_2}}:=\frac{1}{\sqrt2}\left(\ket{v_{p,k}^{s_1s_2}}\ket{k} -
 \ket{v_{p,-k}^{s_2s_1}}  \ket{-k}\right) \\
  &\ket{w_{p,k}^{s_1s_2}}:=\frac{1}{\sqrt2}\left(\ket{v_{p,k}^{s_1s_2}}-
 \ket{v_{p,-k}^{s_2s_1}} \right)
.
\end{align}
By applying the residue theorem we have that
\begin{align}
  \label{eq:Stheintegral}
 &\lim_{\epsilon \to 0^+} \Gamma(e^{-i \omega +
   \epsilon}) = 
   \sum_{(k,s,s') \in \mathsf{R}(\omega)}
  \frac{1}{\partial_k\omega_{p,k}^{ss'} }
   P^{ss'}_{p,k} + R\\
  \nonumber
  &(k,s,s') \in \mathsf{R}(\omega) \iff
    \begin{cases}
       e^{-i \omega}=e^{-i\omega_{p,k}^{ss'}}  \\
       \sin(2k) \geq 0  & \mbox{if } \omega \geq 0  \\
       \sin(2k) < 0  & \mbox{if } \omega < 0 
     \end{cases}\\
    \nonumber
  & P^{ss'}_{p,k} :=
    \ketbra{u^{s}_{p+k}}{u^{s}_{p+k}} \otimes\ketbra{u^{s'}_{p-k}}{u^{s'}_{p-k}}\\
    \nonumber
  &R =
  \begin{pmatrix}
    (e^{-i h_+} -1 )^{-1}&0 &0 &0 \\
    0 &0 &0 &0 \\
    0&0 &-1 &0 \\
    0&0 &0 &(1-e^{-ih_-})^{-1} 
  \end{pmatrix}\\
    \nonumber
&  h_\pm := \omega_{p,k}^{s_1s_2} \pm 2p.
\end{align}
For example, let us study the improper matrix elements
$\bra{\Psi_{p,k}^{++}}  S-I
  \ket{\Psi_{p',k'}^{++}}$
with $ 0 \leq k,k' \leq \pi/2$.
By explicit computation one verify that
$ 0 \leq \omega_{p,k}^{++}\leq \pi$ and
$\mathsf{R}(\omega_{p,k}^{++}) =\{ (k,+,+), (k-\pi,-,-)\}$.
Moreover, we have that
\begin{align}
  \label{eq:Snotazionex}
  \begin{aligned}
  &  \ket{w_{p,k}^{++} }=
  \frac{    
  x_{p,k}-y_{p,k}}{\sqrt2}
    \begin{pmatrix}
      0 \\
      1\\
      -1 \\
      0
    \end{pmatrix} \\
&x_{p,k} := \alpha_{+,\uparrow}(p+k) \alpha_{+,\downarrow}(p-k ), \\
&y_{p,k} :=    \alpha_{+,\downarrow}(p+k) \alpha_{+,\uparrow}(p-k).
  \end{aligned}
\end{align}
Then, we restrict $\Gamma$ to the two
dimensional subspace
corresponding to the support of the projector
$Q$ which is defined as follows: $Q: (a,b,c,d) \mapsto (0,b,c,0)$. We obtain
\begin{align}
  \begin{aligned}
      \lim_{\epsilon \to 0^+ } &\Gamma(\exp(-i
  \omega_{p,k}^{++})) = \\
 & = \frac{x_{p,k}}{   y^2_{p,k} - x^2_{p,k}  }
   \begin{pmatrix}
    x_{p,k}&   y_{p,k} \\
y_{p,k} &x_{p,k}
   \end{pmatrix},
  \end{aligned}
\end{align}
from which we have:
\begin{align}
  \label{eq:Sfirstorder}
  & \begin{aligned}
   &  \bra{\Psi_{p,k'}^{++}}    S-I
\ket{\Psi_{p,k}^{++}} = \delta(k-k') \cdot\\
  &\cdot \frac12 \frac{y_{p,k} -  x_{p,k}}{y_{p,k}+x_{p,k}}  \left( \lambda 
                                  - \frac{2x_{p,k}}{x_{p,k}+y_{p,k}}\lambda^2
                                  + o(\lambda^2)\right)   ,
                              \end{aligned}\\
   &  \bra{\Psi_{p,k-\pi}^{--}}    S-I
\ket{\Psi_{p,k}^{++}} = - \bra{\Psi_{p,k}^{++}}    S-I
     \ket{\Psi_{p,k}^{++}} ,\\
   &  \bra{\Psi_{p,k-\pi}^{--}}    S-I
\ket{\Psi_{p,k}^{--}} =  \bra{\Psi_{p,k}^{++}}    S-I
     \ket{\Psi_{p,k}^{++}} .
\end{align}
Moreover, we can sum the  Born series and obtain the following closed
form for the $T$ matrix 
\begin{align}
  \label{eq:SfullTmatrix}
  \begin{aligned}
   \lim_{\epsilon \to 0^+ } T(\exp(-i
                              \omega_{p,k}^{++}))  
  = \frac{\lambda}{(\lambda+1)^2 x_{p,k}^2 - y_{p,k}^2} \\
   \hphantom{=} \cdot 
  \begin{pmatrix}
    (\lambda+1) x_{p,k}^2 - y_{p,k}^2 & -\lambda x_{p,k} y_{p,k} \\
    -\lambda x_{p,k} y_{p,k} & (\lambda+1) x_{p,k}^2 - y_{p,k}^2   
  \end{pmatrix},
\end{aligned}
\end{align}
from which we have
\begin{align}
  \label{eq:Sfullamplitude}
  \begin{aligned}
      &\bra{\Psi_{p,k'}^{++}}    S-I
      \ket{\Psi_{p,k}^{++}} = \delta(k-k') \cdot \\
& \cdot      \frac{\lambda (y_{p,k} - x_{p,k})
    }{2((\lambda+1)x_{p,k}+y_{p,k})},
  \end{aligned}
\end{align}
which is in accordance with the Bethe ansatz diagonalization of Ref.\cite{PhysRevA.97.032132}.


\subsection{Dyson series}

The scattering amplitude of the Thirring Quantum Cellular
Automata
can be perturbatively calculated by using the Dyson series
for discrete time dynamics of Equation
\eqref{eq:powerexpSmatrix}
in the main text.

In the interaction picture, the interacting Hamiltonian
reads as follows:
\begin{align}
  \begin{aligned}
     H_I(t) &:= \chi \sum_{x \in \mathbb{Z}}
  \psi^\dag_{\uparrow}(x,t) \psi_{\uparrow}(x,t)
  \psi^\dag_{\downarrow}(x,t)\psi_{\downarrow}(x,t), \\
  \psi_{a}(x,t) &= \\
  =\int_{-\pi}^\pi&
  \frac{dk}{\sqrt{2\pi}}\sum_{s=\pm} \alpha_{s,a}(k)
  \psi_s(k) e^{-i(\omega(k)t+kx)}.
  \end{aligned}
\end{align}


The contraction of a pair of fermionic field operators
is given by the difference between the time ordered product
and the normal ordered product,
for example
\begin{align}
  \begin{aligned}
    \contraction{}{\psi_a}{(x,t)}{\psi^\dag_b}
    \psi&_a(x,t)\psi^\dag_b(x',t') := \\
    &=\mathsf{T}[\psi_a(x,t)\psi^\dag_b(x',t') ] -
    \mathsf{N}[\psi_a(x,t)\psi^\dag_b(x',t')],  \\
    \mathsf{T}&[\psi_a(x,t)\psi^\dag_b(x',t') ]  := \\
    &\qquad \qquad\theta(t-t') \psi_a(x,t)\psi^\dag_b(x',t') + \\
    &\qquad \qquad - (\theta(t'-t)-\delta_{t,t'})
    \psi^\dag_b(x',t')
    \psi_a(x,t), \\
    \mathsf{N}&[\psi_a(x,t)\psi^\dag_b(x',t')] :=
    -\psi^\dag_b(x',t') \psi_a(x,t),\\
    \theta&(t) :=
  \begin{cases}
    1 & t \geq0 \\
    0 & t<0
  \end{cases},
        \end{aligned}
\end{align}
and analogously for the products $\psi_a(x,t)\psi_b(x',t')$,
$\psi^\dag_a(x,t)\psi_b(x',t')$ and
$\psi^\dag_a(x,t)\psi^\dag_b(x',t')$.
The appearence of the retarded propagator is a consequence
of the pseudo-particle representation \eqref{eq:S25}.
Moreover, since we are considering a discrete theory,
the product of field operators at the same time and location
 are well defined and must be taken into account in the
 calculation. 
 By a straightforward calculation we have
 \begin{align}
   \begin{aligned}
      &\begin{aligned}
   \contraction{}{\psi_a}{(x,t)}{\psi^\dag_b}
  \psi_a(x,t)&\psi^\dag_b(x',t') =\\
&=\theta(t-t')\int^{\pi}_{-\pi}\frac{dk}{2 \pi
  }\sum_{s=\pm}    \ketbra{u^{s}_{k_1}}{u^{s}_{k_1}}_{ab}
  \cdot \\
 &\quad \cdot e^{-i(s\omega(k) (t-t') + k (x-x'))}      
\end{aligned}\\
&   \begin{aligned}
     \contraction{}{\psi^\dag_b}{(x',t')}{\psi_a}
                     \psi^\dag_b(x',t')\psi_a(x,t) =& -
      \contraction{}{\psi_a}{(x,t)}{\psi^\dag_b}
                     \psi_a(x,t)\psi^\dag_b(x',t')   + \\
 &  +\delta_{t,t'}\delta_{x,x'} \delta_{a,b}             
\end{aligned}\\
&   \contraction{}{\psi_b}{(x',t')}{\psi_a}
                  \psi_b(x',t')\psi_a(x,t)=
                  \contraction{}{\psi^\dag_b}{(x',t')}{\psi^\dag_a}
                     \psi^\dag_b(x',t')\psi^\dag_a(x,t) =0. 
   \end{aligned}  
 \end{align}
It is also convenient to define the following external leg
contractions:
\begin{align}
  \begin{aligned}
 &  \contraction{}{\psi_s}{(k)}{\psi^\dag_a}
  \psi_s(k)\psi^\dag_a(x,t)=
  \frac{{\alpha}_{s,a}(k)}{\sqrt{2}}e^{i(s\omega(k)t+kx)} \\
 & \contraction{}{\psi_a}{(x,t)}{\psi^\dag_s}
 \psi_a(x,t)\psi^\dag_s(k)=
  \frac{{\alpha}_{s,a}(k)}{\sqrt{2}}e^{-i(s\omega(k)t+kx)}
  \\
 & \contraction{}{\psi^\dag_s}{(k)}{\psi_a}
  \psi^\dag_s(k)\psi_a(x,t)=
 \contraction{}{\psi^\dag_a}{(x,t)}{\psi_s}
 \psi^\dag_a(x,t)\psi_s(k)=
  0.
  \end{aligned}
\end{align}
It is now possible to apply the Wick theorem
and compute the terms of the Dyson series.

\subsubsection{First order}
Let us consider the scattering of  two particles.
The first order term in the Dyson series reads as follows:
\begin{align}
  \label{eq:Streelevel7}
  \begin{aligned}
    \sum_{t \in \mathbb{Z}} \bra{\Omega} \psi_{s_4}(k_4) \psi_{s_3}(k_3)
H_I(t)               \psi_{s_1}(k_1)
  \psi_{s_2}(k_2)\ket{\Omega} =\\
  = \sum_{x,t \in \mathbb{Z}} \bra{\Omega} \psi_{s_4}(k_4) \psi_{s_3}(k_3)
  \psi^\dag_{\uparrow}(x,t) \psi_{\uparrow}(x,t) \cdot  \\
               \cdot \psi^\dag_{\downarrow}(x,t)\psi_{\downarrow}(x,t)
               \psi_{s_1}(k_1) \psi_{s_2}(k_2)\ket{\Omega}.
  \end{aligned}
\end{align}
Contracting two field of the same interaction term gives
zero and we do not have any higher order correction
to the free propagation, as expected.
Then, a fully contracted term in Equation
\eqref{eq:Streelevel7} corresponds to the diagram
\begin{align}
  \vcenter{\hbox{
  \includegraphics[width=0.0625\textwidth]{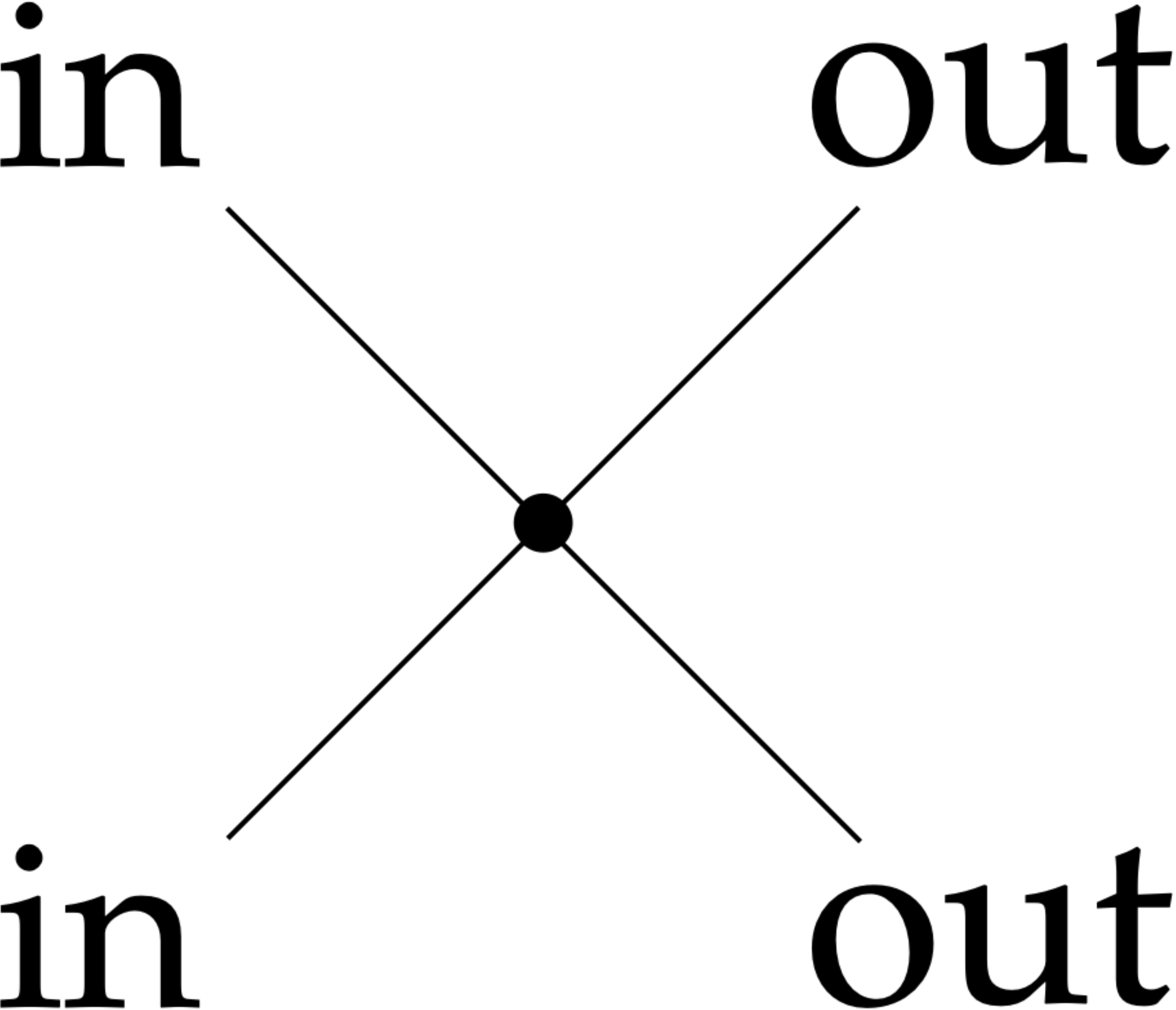}}}
  \;\; .
\end{align}
We have four contribution of this kind which gives the
following amplitude:
\begin{align}
  \label{eq:Sscatteringtree2}
 & \begin{aligned}
  \bra{\Omega}& \psi_{s_4}(k_4) \psi_{s_3}(k_3)
(S-I)           \psi_{s_1}(k_1)
  \psi_{s_2}(k_2)\ket{\Omega}  =\\
 &= \delta_{2\pi} (\omega - \omega')
  \delta_{2\pi} (p - p') 
  \braket{w_{p,k'}^{s'_4s'_3}}{w_{p,k}^{s_4s_3}} =
\end{aligned}\\
\nonumber
  &\begin{aligned}
  \omega := s_1\omega(k_1)  + s_2 \omega(k_2),
  \\
  \omega' := s_3\omega(k_3)  + s_4 \omega(k_4)\\
  p := \frac{k_1+k_2}{2} \mod 2\pi,\\
  p' := \frac{k_3+k_4}{2} \mod 2\pi,\\
  k := \frac{k_1-k_2}{2} \mod 2\pi,\\
  k' := \frac{k_3-k_4}{2} \mod 2\pi.
  \end{aligned}
\end{align}
From Equation~\eqref{eq:Sscatteringtree2}  we obtain,
for example:
\begin{align}
  \bra{p+k',+&; p-k',+} S-I \ket{p+k,+; p-k,+} =\\
  &=\delta(k-k') i\chi \frac12 \frac{y_{p,k} -
    x_{p,k}}{y_{p,k}+x_{p,k}} + o(\chi^2)
\end{align}
which,correctly
coincides with the leading order term in the expansion
of Equation \eqref{eq:Sfirstorder}.

\subsubsection{Second order}
The second order term in the Dyson series is: 
\begin{align}
  \label{eq:Scaramel7}
  \begin{aligned}
      \sum_{\substack {x,x' \in
  \mathbb{Z}\\  t, t'}}&
  \bra{k_4s_4,k_3 s_3}
\mathsf{T}[H_{I}(t,x) H_{I}(t',x')]
  \ket{k_2,s_2;k_1, s_1} =\\
 =& \sum_{\substack {x,x' \in
  \mathbb{Z}\\  t, t'}}
  \bra{k_4,s_4;k_3, s_3}
  \mathsf{T}[\psi^\dag_{\uparrow}(x,t) \psi_{\uparrow}(x,t) \cdot\\
& \quad \cdot      \psi^\dag_{\downarrow}(x,t)\psi_{\downarrow}(x,t)
                                                           \psi^\dag_{\uparrow}(x',t') \psi_{\uparrow}(x',t') \cdot\\
 &\quad \cdot  \psi^\dag_{\downarrow}(x',t')\psi_{\downarrow}(x',t')] \ket{k_2s_2,k_1 s_1}
  \end{aligned}
\end{align}
and a fully contracted term corresponds to the following
diagram:
\begin{align}
  \vcenter{\hbox{
  \includegraphics[width=0.125\textwidth]{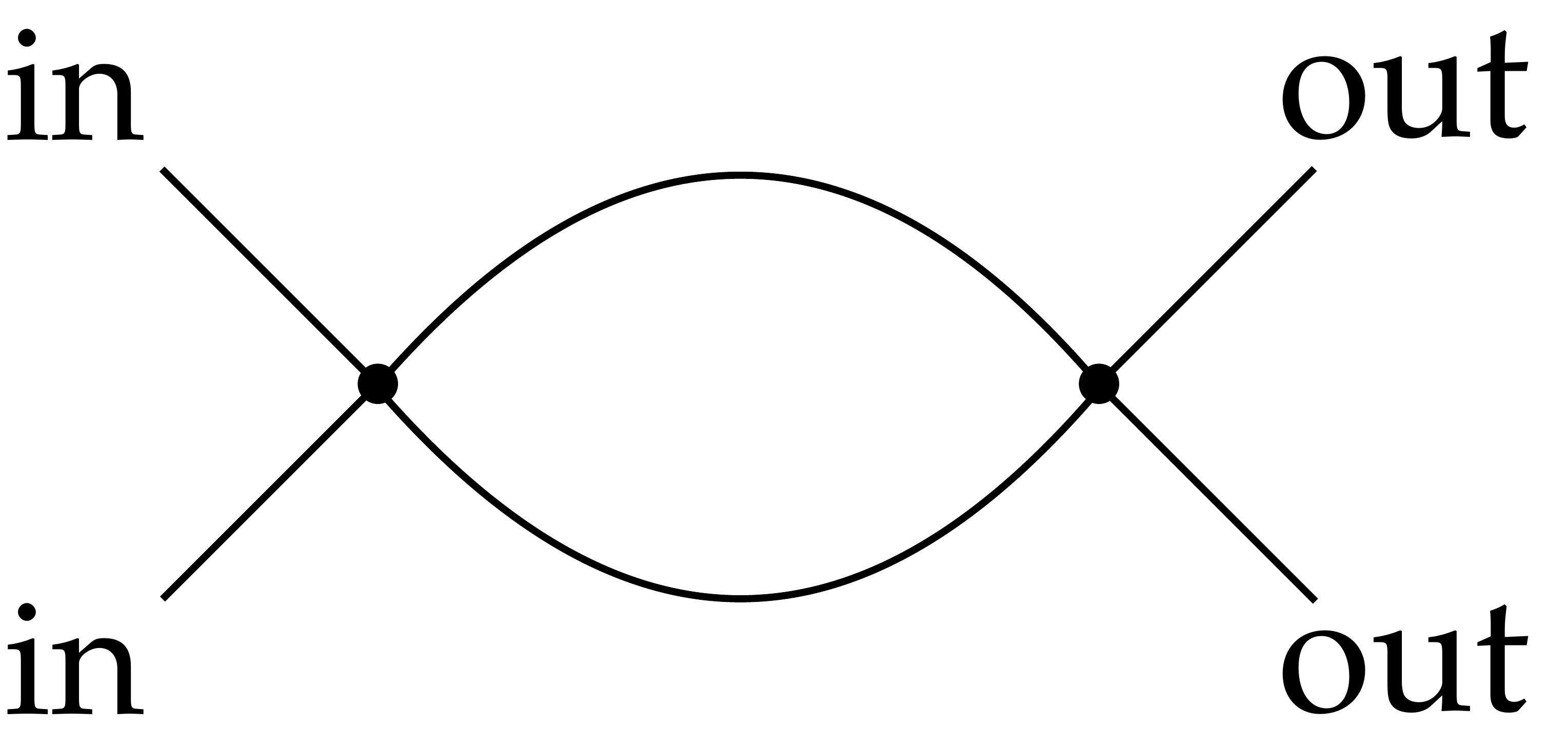}}}
  \;\; .
\end{align}
A lenghty but straightforward calculation leads to the
following result
\begin{align}
  \label{eq:Sperturbation3}
  &\bra{p+k',+; p-k',+} S-I \ket{p+k,+; p-k,+} =\\
            &
              =  \delta(k-k') \frac12 \frac{y_{p,k} -
              x_{p,k}}{y_{p,k}+x_{p,k}}
              \bigg(
              i\chi + \\
  &\left.\quad
              +\left( \frac12 -
  \frac{2x_{p,k}}{x_{p,k}+y_{p,k}} \right)
  (i \chi)^2
              + o(\lambda^2)\right),
\end{align}
which agrees with Equation~\eqref{eq:Sfirstorder}

The $T$ operator is given by the following finite
dimensional matrix
\begin{align}
  \label{eq:18}
  \begin{aligned}
&      T(z) = \lambda {C}+\lambda^2{C}
(z-D_p^{(2)})^{-1}D_p^{(2)}
  {C}, \\
 & {C} := I \otimes \ketbra{0}{0}, \quad \lambda :=  e^{i \chi}-1.
  \end{aligned}
\end{align}
Since $C$ is a projector, from Equation~\eqref{eq:18} we
have that the series~\eqref{eq:12} and
\eqref{eq:13}  become the following geometric series:
\begin{align}
  \label{eq:21}
  \begin{aligned}
      &\bra{k'}\bra{v_{k'}^{r's'}}S-I \ket{v_{k}^{rs}}\ket{k} = 
     \delta_{2 \pi} (\omega_{s'r'}(k')-\omega_{sr}(k) )\cdot \\
   &  \lim_{\epsilon \to 0^+}\bra{v_{k'}^{r's'}} \left( 
\sum_{n=0}^{+\infty}
\lambda^{n+1} \Gamma^n_0(e^{-i \omega_{sr}(k)  + \epsilon})  \right)\ket{v_{k}^{rs}}, \\
&\Gamma_0(z) :=  \int_{-\pi}^{\pi}\sum _{a,b = \pm} \frac {e^{-i
    \omega_{ab}(s)} } {z -  e^{-i\omega_{ab}(s)}}\ketbra{v_{s}^{ab}}{v_{s}^{ab}} \,ds
  \end{aligned}
\end{align}

\end{document}